\documentclass[letterpaper,11pt,onecolumn,accepted=2022-05-11]{quantumarticle}
\pdfoutput=1

\usepackage[utf8]{inputenc}
\usepackage[english]{babel}
\usepackage[T1]{fontenc}
\usepackage{amsmath,amsthm,amsfonts,bm}
\usepackage{csquotes}
\usepackage{comment,float}
\usepackage[margin=1in]{geometry}
\usepackage[colorlinks]{hyperref}
\usepackage{graphicx}
\usepackage[backend=biber,sorting=none,style=ieee]{biblatex}
\graphicspath{ {./images/} }
\usepackage{physics}

\usepackage{tcolorbox}

\begin{document}

\title{Quantum and Classical Bayesian Agents}
\author{John B. DeBrota and Peter J. Love}
\affiliation{Department of Physics and Astronomy, Tufts University, Medford, MA 02155, USA}

\maketitle

\begin{abstract}
    We describe a general approach to modeling rational decision-making agents who adopt either quantum or classical mechanics based on the Quantum Bayesian (QBist) approach to quantum theory. With the additional ingredient of a scheme by which the properties of one agent may influence another, we arrive at a flexible framework for treating multiple interacting quantum and classical Bayesian agents. We present simulations in several settings to illustrate our construction: quantum and classical agents receiving signals from an exogenous source, two interacting classical agents, two interacting quantum agents, and interactions between classical and quantum agents. A consistent treatment of multiple interacting users of quantum theory may allow us to properly interpret existing multi-agent protocols and could suggest new approaches in other areas such as quantum algorithm design.
\end{abstract}

\section{Introduction}

Quantum theory is our most successful physical theory. It accurately predicts experimental statistics in the widest range of empirical scenarios. However, it is also our most mysterious physical theory because it is not clear how to connect its instrumental success to a physical conception of reality. Quantum theory appears to be ``right,'' but, after nearly a century, we still lack an explanation for why. What makes quantum theory the right theory?

As long as it is clear how to use quantum theory, one might defer this question. Is it always clear how to apply quantum theory to an arbitrary situation? A famous suggestion that it is not is given by the ``Wigner's friend'' thought experiment \cite{Wigner1961}. Wigner's friend will make a quantum measurement of a system in a lab and obtain an outcome. Wigner himself is outside the lab and quantum mechanically treats the whole lab containing his friend and the system. For Wigner, the joint system evolves continuously in time; in particular, there is no outcome of the measurement in the lab. Wigner and his friend both use quantum theory correctly, but arrive at contradictory conclusions. Who is right? There remains no universal consensus on a resolution.

Recent extensions of the Wigner's friend thought experiment probe more deeply into the lack of complete consensus in regimes consisting of multiple interacting agents. Reference \cite{Frauchiger2018} features a thought experiment involving a more elaborate setup than the original Wigner's friend scenario. At the end of a sequence of events, two agents who both seem to correctly employ quantum mechanics, appear to make contradictory predictions about the outcome of a measurement Wigner will make. The setup in reference \cite{Baumann2020} is conceptually simpler, but still features two agents forecasting the outcome of a measurement Wigner will make. In this case, not only is there an apparent contradiction, but the authors suggest that the friend is actually misled by unmodified quantum theory. As noted in both works, what one makes of these thought experiments hinges on one's interpretation of quantum theory. Do these point to a breakdown in the applicability of quantum theory? 

One might also ask: what is an agent? In the works above, agent generically means user of quantum theory. This too relies on one's interpretation. Depending on what one thinks quantum theory \emph{is}, it is potentially a different thing to use it. For example, Bohr's emphasis on the necessity of classical descriptions of experiments and data may inspire a conception of user of quantum theory somehow based on classical mechanics. Furthermore, one wonders whether invoking the full apparatus of quantum theory in a multi-agent setting is appropriate without a formal treatment of agents. At the very least, there is clearly confusion about how to fit together quantum theory and agency.

This confusion could have practical consequences. Demonstrations of the informational power of quantum theory often presume a multi-agent setting. For example, multi-prover quantum games, such as XOR games like the Peres--Mermin Magic Square \cite{Peres1990,Mermin1990}, employ two or more agents who can share resources like quantum entanglement. Quantum versions of Merlin--Arthur protocols \cite{Kitaev2002,Vidick2013} require transmission of quantum information between differently capable agents: Merlin, a wizard, can magically produce quantum proofs and send them to Arthur, a human, who might hope to verify the proofs with the aid of a quantum computer. Another potential multi-agent scenario is information theoretically secure quantum voting~\cite{Arapinis2021,Khabiboulline2021}. When multiple agents are involved, phenomena related to their coordination, cooperation, and even competition may become nontrivial. It must be understood what it means to be an agent capable of performing quantum information processing tasks in order to properly interpret their consequences. Quantum technology may be held back by conceptual confusion.

To address these concerns, we turn to the Quantum Bayesian, or QBist, foundational perspective on quantum theory. More so than any other interpretation, QBism brings agency into the foreground. A QBist regards quantum theory as a tool that a single agent may adopt to help manage their expectations for the consequences of their actions. For QBism, the basic foundational question we noted in the first paragraph becomes: In what kind of universe are agents well-advised to adopt the quantum formalism to make better decisions? The goal is to learn something about physical reality from the shape of our best means for navigating it. There are many expositions on QBism~\cite{Fuchs2016a,Fuchs2010,Fuchs2013}. In the interest of making the present paper self contained, we will introduce the aspects of QBism we rely on and refer the reader to these references for further background. 

Quantum theory, on this account, is \emph{for} agents, and, as such, does not directly \emph{define} them; agency is undefined at this level of treatment. This is not a claim that agency is physically fundamental (nor a claim that it isn't), it is rather the simple point that there must be someone who \emph{uses} a tool. QBism is clear, however, that an agent is fundamentally a decision maker, and where decisions are concerned, the freedom of the agent is paramount. The freedom of action of an agent must never be compromised, not even when another agent assigns a wavefunction to the first; if some process revokes this freedom, the system is no longer an agent. As one of us and collaborators wrote in a previous paper: ``Agents are entities that can take actions on parts of the world external to themselves, so that the consequences of their actions matter to them.''~\cite{DeBrota2020}. 

QBism is the only extant interpretation with a primal role for agency and, as such, is uniquely situated to unambiguously treat the multi-agent setting. Indeed, QBism handles the original and extended Wigner's friend thought experiments without strain: there is no breakdown in the applicability of quantum theory~\cite{DeBrota2020}. Quantum theory is a means for an agent to manage \emph{their} experiences; as far as the formalism is concerned, measurement outcomes are personal to the agent taking actions and so differing accounts \emph{between} agents regarding the occurrence of an outcome is no more unusual than the fact that different people have different experiences. Paradoxes arise when one agent's use of quantum theory is assumed to pertain to another. In \cite{Baumann2020}, it is assumed the friend should adopt Wigner's wavefunction for their reasoning on a system including themselves, which amounts to robbing them of their agency. Removing inconsistent conclusions from \cite{Frauchiger2018} similarly reduces to denying the universality of an assumption of agreement between agents. When Wigner and all of his associates are appropriately granted equal and symmetric status as users of quantum theory, no paradoxical conclusions may be reached. 

Although we don't need to model ourselves to use quantum theory, we do need some sort of agent model to simulate multiple users of it. In this paper, we propose a QBist approach to modeling multiple interacting agents who utilize quantum or classical mechanics. To accomplish this, we focus on a particular kind abstract agent defined entirely in terms of what they do and then endow these agents with quantum or classical reasoning. We feign no physical mechanism for agency. Rather, we model agents as instances of a few structures that an actual agent wishing to take particular kinds of quantitatively informed actions would use for managing their own decisions. It should be clear that we are not aspiring to anything like a ``realistic'' model of agency. 

Specifically, we model so-called ``rational'' agents, which are idealized Bayesian decision-makers \cite{Bernardo2000} who always act by maximizing their expected utility \cite{Schoemaker1980}. Rational agents consistently use probability theory, a prerequisite for consistent use of quantum theory. However, there is nothing obviously physical, neither classical nor quantum, about probability theory or its conceptual parent, decision theory. Thus, we require an additional ingredient to endow rational agents with classical or quantum mechanics. 

This ingredient is the QBist notion of a reference action, or reference measurement. Following QBism, we indifferently use the terms ``action'' and ``measurement'', preferring the former for its lack of passive connotation. We want our model to more or less indifferently work with any physics, that is, we want to be able to equally well model an agent who adopts quantum mechanics, classical mechanics, or potentially something else altogether. We call the relation between one's probabilities for the outcomes of a reference action and the probabilities for the outcomes of any other action a \emph{physical postulate}. As we will see, a ``classical'' addition to probability theory is one such postulate, while the Born rule is another. This framework permits other physical postulates to be considered as well, but here we will focus on the two most relevant: those associated with quantum and classical mechanics. In order to usefully wield a physical postulate, our agents must possess the ability to take some number of distinct actions and to receive outcomes, to choose between such actions, and to learn from their outcomes. The rational agent paradigm formalizes these latter properties. Thus, our approach is what results when rational agents adopt a physical postulate. 

The content of the paper consists of two broad parts: First, in \S\ref{sec:rationalagents}, \S\ref{sec:refactions}, and \S\ref{sec:inference}, we introduce and formalize the atomic element of QBist agent-based modeling, a single agent, by reviewing Bayesian rational agents, explaining what is needed to elevate them to classical or quantum agents, and showing that such agents can learn in simple tomographic scenarios. Second, in \S\ref{sec:interaction}, we introduce the idea of an interaction, the remaining piece needed to model \emph{multiple} agents, and  propose an interaction scheme called expectation sampling to illustrate our model in the simplest case of two interacting agents in a variety of settings. A reader who is sufficiently familiar with QBism and state tomography by Bayesian mean estimation may treat the first part mostly as a reference. We also note that our simulations and analysis in the second part are primarily illustrative. A fully-fledged numerical study will appear in a future work.

More specifically, the rest of the paper is organized as follows. In \S\ref{sec:rationalagents}, we introduce rational agents within the context of Bayesian decision theory. In \S\ref{sec:refactions}, we present the idea of a reference action from QBism and, through it, define the notion of a physical postulate that a particular agent adopts. In \S\ref{sec:postulates}, this idea is formalized: A classical agent adopts a trivial postulate structurally equivalent to the law of total probability while a quantum agent adopts a postulate derived from the Born rule given a particular reference action. In \S\ref{sec:inference} we turn to the topic of agents learning from experience. In \S\ref{sec:infexch}, it is seen that rational agents who believe the outcomes of a sequence of repeated actions are \emph{exchangeable}, a generalization of independent and identically distributed outcomes, may engage in parameter estimation; to illustrate this, we simulate a rational agent arriving at an estimate for the bias of a coin on the basis of many repeated flips. With the addition of a physical postulate, agents with exchangeable prior beliefs possess sufficient sophistication to do something even more like science. In \S\ref{sec:infwithpostulates}, we show that with an exchangeable joint reference prior, a quantum or classical agent may implement state tomography of systems received from an exogenous source where they are not restricted to repeating the same action every time. We illustrate this in \S\ref{sec:tomography} for the case of a quantum agent who receives qubits prepared by an exogenous source and randomly takes one of the three Pauli measurements on each system. In \S\ref{sec:interaction}, we introduce the notion of an interaction between two agents which we heuristically motivate in terms of broadcasting and receiving signals. In \S\ref{sec:priorsampling}, out of the many possible choices, we propose an interaction we call expectation sampling in which the expectation of the prior of one agent distributes the outcomes for the other. In the remaining subsections, we illustrate our framework with expectation sampling by simulating two interacting classical agents in \S\ref{sec:bernoulli}, two interacting quantum agents in \S\ref{sec:2qubit}, and interactions between a classical and a quantum agent in \S\ref{sec:qc}. In \S\ref{sec:disc}, we discuss many possible next steps and conclude.

\section{Rational agents}\label{sec:rationalagents}

A proper use of quantum theory presumes a proper use of probabilities. Notwithstanding centuries of debate over the meaning of probability theory, it is fairly unobjectionable to say that the \emph{use} of a probability is to inform action. If the weather forecast says there is a 70\% chance of rain, we may \emph{choose} to take an umbrella. It is therefore pleasing to discover that one may found probability theory itself in a theory of decision making. Specifically, probabilities naturally arise in the course of developing so-called subjective Bayesian decision theory. Bayesian decision theory addresses how agents who strive to meet a norm of behavioral self-consistency called quantitative coherence should act in the face of uncertainty~\cite{Bernardo2000}. As we already wished to address agency, the fact it organically appears within this framework indicates this is the right place for us to start.

Quantitative coherence, hereafter simply coherence, axiomatizes the idea that one should strive to act in a way consistent with one's own interests or preferences. In fact, there have been numerous inequivalent formulations of the principles of coherence resulting in the same decision criterion. Here we are interested in applying the resulting theory, so we are not concerned with the nuances of these differences. The interested reader may see \cite{Fishburn1981} for a review. The common intuitive thread is this: If one's preferences meet certain criteria, coherence implies that an agent's quantified degrees of belief are probabilities, that their preferences among the consequences of their actions are assigned numerical values by a utility function, and that their most preferred actions are those which maximize their expected utility. 

One may get the basic idea of coherence as it pertains to probabilities from the ``Dutch book'' story~\cite{deFinetti1937}. Here dollars are loosely considered a valuation of one's preferences; one is assumed to always prefer to gain money rather than lose it. Consider a simple gambling scenario in which someone is contemplating the purchase and/or sale of lottery tickets which pay \$1 if an event occurs and \$0 if it does not. With respect to a given event, an agent may regard a ticket to be worth a certain amount of money, implying that they would equally prefer to buy or sell it at that price. If an agent regards the prices for a set of tickets to be fair in this way and yet there exists some combination of buying and/or selling the tickets that is guaranteed to lead to a net loss of money for the agent, irrespective of which events occur, we say that set of prices is \emph{incoherent}. A set of prices is coherent if it is not incoherent. 

This story provides an operational foundation to probability theory itself: An agent's prices for the tickets associated with a set of events are coherent if and only if the numerical value of those prices are probabilities satisfying the usual rules of probability theory. In particular, it may be shown that coherence implies for any event $E$ that $0 \leq p(E) \leq 1$,
that $p(E) = 1$ if the agent believes $E$ is certain to occur, and that $p(E \vee D) = p(E) + p(D)$ if the agent believes $E$ and $D$ are mutually exclusive. Conditional probabilities are similarly defined in terms of a conditional lottery, i.e.,\ lotteries where transactions are canceled if the event being conditioned upon does not occur. Coherence then implies the relation that for any events $A$ and $B$, $p(A|B)=p(A,B)/p(B)$.

We define an event for a particular agent as the consequence, or outcome, of an action they took. Typically an event is thought to be a fact in the world, verifiable in principle by anyone, while a consequence is what that outcome means to the agent, but we unify these concepts in anticipation of quantum theory where measurement outcomes cannot meet this standard of objectivity; for consistency, one is led instead to the notion of ``facts for the agent'' \cite{Caves2007}. We use the word ``action'' to refer to anything an agent might do, for example, pressing a button, listening attentively, or measuring the spin polarization of an electron. An agent's probabilities, which they might arrive at by imagining lotteries of the type described above, are then numerical valuations of their relative likelihoods for the outcomes of any action they may take. 

The basic decision problem for an agent is to determine which of a number of potential actions they most prefer given their preferences among outcomes and their beliefs about the relative likelihood of such outcomes. As mentioned above, coherence implies the use of probabilities to quantify an agent's degree of belief for the relative likelihood for the consequences of their actions. Coherence also implies that the agent's relative preferences among the potential outcomes of actions may be quantified with a utility function. A utility function is a real-valued function over the set of consequences encoding an abstraction of the benefit, or utility, that a given outcome would grant to the agent. In the Dutch book story, the use of dollars amounts to a simple utility function where the benefit granted by a single dollar is independent of one's wealth. Actual utility functions may be much more complex.

In non-Bayesian approaches to decision theory, probability and utility are often considered separately --- a theory of utility is built upon a freestanding notion of probability. However, in the coherentist paradigm, they are developed in tandem, both following from the norm of coherence. With probabilities and utilities in hand, the decision problem may be solved: it may be shown that coherence implies an agent most prefers actions which maximize their expected utility~\cite{Bernardo2000}. 

Coherence codifies a kind of normative rationality. It is ``normative'' in the sense that it is imagined one \emph{should} strive to avoid incoherence and its consequences. Supposing one accepts this norm doesn't, of course, mean that one always meets it. One may lack sufficient resources, for example, time, money, or computing power, to ensure perfect consistency. However, to distinguish between the consequences of failing to meet a norm and the consequences of the norm itself, it is interesting to explore the idealized situation of agents who always meet the norm. Towards this, we define a \emph{rational agent} to be an idealized agent who is always coherent. We consider rational agents because they always use probability theory properly and so they stand a chance of using quantum theory properly as well. Going forward, we restrict attention to this kind of agent.

To illustrate how a rational agent behaves, consider the following example. Suppose a rational agent, Alice, faces a decision between three actions $A$, $B$, and $C$ which have outcomes $a\in \{1,\ldots,N_A\}$, $b\in \{1,\ldots,N_B\}$, and $c\in \{1,\ldots,N_C\}$. Alice represents her preferences among these outcomes with a utility function $u(x)$ which takes real values for all $a$, $b$, and $c$. She also represents her beliefs in the relative likelihood of the outcomes with probability distributions $p_A(a)$, $p_B(b)$, and $p_C(c)$. Her expected utility for the action $A$ is 
\begin{equation}
    \sum_{a=1}^{N_A} u(a)p_A(a)\;,
\end{equation}
and her expected utilities for $B$ and $C$ are calculated similarly. The rational choice for Alice is then to take the action for which she has the largest expected utility. In case more than one action equally maximizes her expected utility, any of the maximizing actions is rational.

Finally, it's important to compare what we have discussed with another scenario our readership is likely to encounter. We have discussed how rational agents choose between actions when they have preferences among the uncertain outcomes. Sometimes, however, we are more interested in an agent's probabilistic forecast itself than the consequences of the actions for which they have probabilities. Such a situation is especially important in science where the goal may be to report a probabilistic summary of what has been learned. So-called ``pure inference scenarios'' may be thought of as a special case of the basic decision problem we have introduced~\cite{Bernardo2000}. In a pure inference scenario, the actions one decides between are ``inference statements'', an example of which could be a probability distribution for another action, to report. The possible consequences from this perspective are the pairs consisting of an inference statement and an outcome of the considered action. One may see how an agent may then prefer some consequences to others --- perhaps they prefer to have asserted a probability distribution which deemed the actual outcome a likely event. If the utility function is an example of a ``proper score function'', then the rational, i.e.\ expected utility maximizing, action for an agent is to report their true beliefs about the likelihood of an event. For an example of this type of situation involving quantum mechanics, see \cite{Blume-Kohout2006}.

In the next section, we will leave the pure decision theoretic setting and our agents will contend with something more like physics.

%Having introduced rational agents, we now address their suitability for a simple scientist model. Perhaps the most important characteristic necessary for this task is an ability to \emph{learn} from experience. In the next subsection, we will see how and in what circumstances rational agents may engage in a particular form of this activity.

\section{Reference Actions}\label{sec:refactions}

Two rational agents, Alice and Bob, are playing the following guessing game. Bob will roll one of three differently colored six-sided dice. Before Bob chooses a die, Alice wishes to predict what number will result from the roll. Alice believes Bob will pick the die with color $j$ with probability $c(j)$ and that the number $i$ will appear given a roll of die $j$ with probability $R(i|j)$. These are Alice's quantified degrees of belief regarding the outcomes of the two actions ``which color'' and ``which number''. Given $R(i|j)$ and $c(j)$, Alice's unconditional probability for ``which number'' is
\begin{equation}\label{ltpdice}
n(i)=\sum_{j=1}^3 R(i|j)c(j)\;.
\end{equation}
This is an application of the Law of Total Probability (LTP) and cannot be violated by a rational agent. 

Consider now the closely related situation where Alice \emph{will not} perform the action ``which color''. What now is Alice's probability for ``which number''? Should it still be given by the LTP? Intuition tells us yes, but this does not follow because Alice is rational. The terms in \eqref{ltpdice} only have meaning when ``which color'' is performed; when it is not, probability theory alone provides no systematic guidance for assigning a probability to ``which number''. To provide justification for using the probabilities from the LTP, an additional assumption is needed. In this case, it is to assume that Alice believes the die Bob rolls has a color whether or not she takes the necessary action to obtain a value. Provided she believes the color of a die is an intrinsic attribute, present irrespective of and prior to its measurement, Alice's expectations for this situation are fully determined by what she would believe were she to take both actions instead.

The sensibility of using the LTP probabilities, which apply in the hypothetical scenario where she takes the action ``which color'', for the actual scenario where the conditioned-upon action is not taken thus follows from a \emph{physical} assumption. The classical interpretation is that measurement is ultimately a passive process which reveals a preexisting fact. Accordingly, whether or not a color measurement is made should make no difference to the probabilities. 

In our example, beliefs about color were relevant for predictions about the number rolled, but no other actions were entertained. It is easy to imagine scenarios where beliefs about the outcomes of a particular action allow us to predict probabilities for a wider range of actual measurements. Remarkably, in classical mechanics, it is actually possible to imagine this wider range encompasses \emph{all} potential measurements! In an idealized classical situation, one could imagine a hypothetical querying of the phase-space coordinates of a system. With a probability distribution over phase space coordinates, also called a Liouville distribution, one may calculate the probability for \emph{any} actual measurement action by performing the appropriate coarse graining, for example, summing up the probabilities for all phase-space coordinates compatible with a particular total energy~\cite{balescu1991equilibrium}. The physical assumption that phase-space coordinates are a preexisting fact of the system and can potentially be passively revealed by measurement again results in probabilities equivalent to those fixed by the LTP for two actual actions, relevant instead to the hypothetical case where the phase-space coordinates are not learned. 

While an agent may believe in passive processes and preexisting facts, the relations between probabilities they imply are not themselves consequences of coherence like the LTP, even when the functional form is the same. The upshot of being mindful of this subtlety is that it is immediately clear that the use of a \emph{different} rule for a hypothetical situation, perhaps due to a different kind of physical assumption, is not a violation of probability theory. Rather, it is an addition. In particular, there is no conflict in allowing rational agents to make physical assumptions or make use of the corresponding rules. In fact, from this perspective, much of the physics is reflected in the rule one uses. Classicality, from this vantage point, corresponds to a rule giving the same probabilities as the LTP. As we will see, quantum theory provides an alternative rule, leaving us to ponder what kind of physical assumptions may compel it.

\subsection{Physical postulates}\label{sec:postulates}

Let us now make explicit the rules we consider. In what follows, we borrow some notation and terminology from \cite{DeBrota2020d}. The substantial assumption we make is that an agent believes any action they could take may be understood through its relation to a hypothetical \emph{reference action}. In particular, it is assumed that their probabilities for directly taking an arbitrary action $\mathcal{D}$ are related to their probabilities for the outcomes of a reference action $\mathcal{S}$ and to their probabilities for the outcomes of $\mathcal{D}$ conditional on an outcome of the action $\mathcal{S}$ in the hypothetical scenario in which $\mathcal{S}$ is also performed. Let $p(i)$ denote their reference probabilities for $\mathcal{S}$, $R(j|i)$ denote their probabilities for $\mathcal{D}$ given an outcome of $\mathcal{S}$, and $q(j)$ denote their probabilities for $\mathcal{D}$ when $\mathcal{S}$ \emph{is not} performed. We introduce the notation $r_j(i):=R(j|i)$ for the $i$th entry in the $j$th row of the conditional probability matrix $R$ and let the vector $r_j=\{r_j(i) : 1,\ldots, N\}$ denote the whole $j$th row. Then the rules we consider are those of the form
\begin{equation}\label{physicalpostulate}
    q(j)=\mathcal{F}(p,r_j)\;,
\end{equation}
for a function $\mathcal{F}$ taking the vectors $p$ and $r_j$ as arguments. We say that an agent who always uses a rule like this has adopted a \emph{physical postulate}. 

The dependence on only the row $r_j$ in a physical postulate is a noncontextuality assumption, as it means the probability for the outcome $j$ for two measurements with different $R$ matrices, but coinciding $j$th rows, must be the same. Provided the agent assigns $p$ and $R$ independently, this assumption further implies $\mathcal{F}$ is bilinear in its arguments~\cite{DeBrota2020d}.

When the physical postulate takes the form of the LTP, as in the dice example and the idealized classical system we considered above, we call it \emph{classical}. The $R$ matrix for taking the classical reference action, i.e.\ when $\mathcal{D}=\mathcal{S}$, must be the $N\times N$ identity matrix, where $N$ is the number of outcomes of the reference action, so that $q=p$. This fits with the classical intuition of measurement passively revealing a property. There is a unique and simple classical reference action because, from the classical perspective, the only thing for a reference action to be is the process which reveals the ``true underlying state of the system''.

%Note that a physical postulate as we have defined it may only be one aspect of one's full probabilistic accounting which is recognizably ``classical'' or otherwise. For example, in addition to postulating the existence of phase-space coordinates, one would also generally adopt a Hamiltonian and evolve their Liouville distribution in time with Hamilton's equations. We are intentionally not fixing further details here because we wish to investigate the consequences of differences from this broader vantage point. When we say ``classical'', we therefore more precisely mean ``classical-like''. 

Quantum theory, on the other hand, does not admit a phase-space or an equivalent as an intrinsic attribute of a system. So-called phase-space quantum theory displays this lack of fit in the appearance of improper, negative probabilities~\cite{Ferrie2008,Ferrie2010}. One should be careful not to interpret this to mean that probability theory is ultimately inadequate once quantum effects are involved. What has gone ``wrong'', from our perspective, is imposing a classical physical postulate. Negativity arises from demanding that the quantum rule for associating probabilities with actions, the Born rule, take a form equivalent to the LTP. 

The Born rule may instead be realized by a different physical postulate relating probabilities. Rather than negativity, we simply find quantum theory is not compatible with a classical physical postulate. Actions in quantum mechanics are quantum measurements. To any measurement $\mathcal{D}$ with $N$ outcomes in a Hilbert space of dimension $d$, one associates a set $\{D_j\}$ of $N$ positive semidefinite operators which sum to the identity. Such a set is called a positive operator-valued measure (POVM) and the operators $D_j$ are called \emph{effects}. Given a quantum state, expressed as a density matrix $\rho$, the outcome probabilities of a general measurement with POVM $\{D_j\}$ are given by the Born rule, 
\begin{equation}\label{standardbornrule}
    q(j)=\tr \rho D_j\;.
\end{equation}
Thus, our goal is to reexpress the Born rule \eqref{standardbornrule} as an instance of a physical postulate \eqref{physicalpostulate}. $\rho$ and $D_j$ must be replaced by a relation between a reference probability vector $p$ and a vector $r_j$ of probabilities for the outcomes of $\mathcal{D}$ given an outcome of the reference action.

A reference action is distinguished from an arbitrary action as follows: an agent must believe that the probabilities for the outcomes of such an action provide all that is needed about a system in order to specify probabilities for the outcomes of any other action. Applied to quantum mechanics, this means that the reference probabilities must contain the same information as the quantum state. 

A candidate quantum reference measurement must span $\mathcal{L}(\mathcal{H}_d)$, the space of Hermitian operators on $\mathcal{H}_d$. Then, because the Born rule is an inner product on $\mathcal{L}(\mathcal{H}_d)$, the Born rule probabilities for such a POVM will be in one-to-one correspondence with the quantum state. The dimension of $\mathcal{L}(\mathcal{H}_d)$ is $d^2$, so the POVM must have at least this many effects. When the effects of a POVM span $\mathcal{L}(\mathcal{H}_d)$, we say the POVM is informationally complete (IC); when an IC POVM has precisely $N=d^2$ effects, it is a minimal IC POVM (MIC). For a recent introduction to MICs, see \cite{DeBrota:2018b}. Infinitely many MICs exist and can be constructed in any finite Hilbert space dimension. While one could consider a reference action with more than $d^2$ outcomes, this implies the use of a larger-than-necessary probability space, a case we find less immediately interesting, so, in this paper, a quantum reference action is a MIC reference action. Supposing the MIC is $\{E_i\}$, the reference probability vector is $p(i)=\tr \rho E_i$. For a fixed reference action, a quantum state and a reference probability vector are completely interchangeable.

In addition to the POVM associated to a measurement, a general action in quantum mechanics requires a specification of the state update, that is, a specification of the quantum state the agent uses for the system after they obtain one of the outcomes. There is an infinite class of updates consistent with a given POVM in terms of sets of completely positive maps~\cite{Nielsen2010,Wilde2017}. Here again, an arbitrary choice won't do for a reference action. Just as $\rho$ in \eqref{standardbornrule} needs to be spanned by the effects of the POVM associated with a reference action, the post-measurement states must span $D_j$ in order to fully characterize an arbitrary measurement as a matrix of probabilities conditional on reference measurement outcomes. In addition to this, we need the post-measurement states to be insensitive to the initial quantum state if the specification of the probabilities for an arbitrary action are to be conditional \emph{only} on the outcomes of the reference action. Conceived of as a quantum channel, our restrictions imply a reference action is an entanglement breaking MIC channel (EBMC) with a linearly independent set of post-measurement states \cite{DeBrota2019}. Supposing the post measurement states are $\{\rho_i\}$, the conditional probability matrix is $R(j|i)=\tr D_j \rho_i$.

We can already see that where our abstraction of classicality is trivial, quantum theory is highly nontrivial. Classicality is associated with a unique reference action, while quantum theory allows \emph{any} EBMC to play this role. Furthermore, quantum theory requires a reference action to have $N=d^2$ outcomes for some integer $d\geq 2$, whereas nothing in our setup prohibits us from considering a classical postulate for any integer $N\geq 2$. 

It was shown in \cite{DeBrota:2018} that given a reference action with MIC $\{E_i\}$ and post-measurement states $\{\rho_i\}$, the Born rule \eqref{standardbornrule} may take the form 
\begin{equation}\label{bornrule}
   q(j)=\sum_{i,k=1}^{d^2}[\Phi]_{ik}R(j|i)p(k)=r_j^T\Phi p\;,
\end{equation}
where $\Phi$ is the inverse of the conditional probability matrix for the reference action itself, that is $[\Phi^{-1}]_{ij}=\tr E_i\rho_j$. In this form, the Born rule is an example of a physical postulate \eqref{physicalpostulate}, differing from the classical postulate by the fact that $\Phi$ is not the identity. For any choice of reference action, $\Phi$ is a column quasistochastic matrix, meaning that its columns are real-valued and sum to unity. The possibility of negative numbers in a quasistochastic matrix distinguishes them from stochastic, or Markov, matrices. In fact, $\Phi$ always has some negative values. The presence of negative numbers in $\Phi$ means that $\Phi p$ is a quasiprobability and the grouping $q(j)=r_j^T(\Phi p)$ places \eqref{bornrule} into the the functional form of the LTP where negativity is inevitable. Another grouping, $q(j)=\left(r_j^T \sqrt{\Phi}\right)\!\left(\sqrt{\Phi} p\right)$, relates MICs to a class of discrete quasiprobability representations which include analogs of Wigner functions on a discrete phase space~\cite{DeBrota2020b}.

Equation \eqref{bornrule} displays the quantum physical postulate given a choice of reference action. In an actual model, one would want to fix a particular reference action choice for each agent and thus a particular form of the Born rule. Regardless of the choice, the Born rule is inherently distinct from the classical physical postulate --- $\Phi$ cannot be made to be the identity within quantum theory. 

The fact that $\Phi$ is not the identity in a quantum postulate results in a reference action dependent restriction on the allowed values of $p$ and $R$ such that $q$ is an actual probability without negative values. Another way to say this is that not all valid probability vectors can take the form $p(i)=\tr \rho E_i$ when $\{E_i\}$ is a MIC. For example, a probability $1$ prediction for any outcome of a MIC reference action is not compatible with any quantum state. Fundamentally, this restriction comes from the fact that a MIC must be an operator basis with the additional restriction that every element is positive semidefinite~\cite{DeBrota:2018b}. Consequently, a reference action defines a proper subset of the reference probability simplex corresponding to quantum states and a proper subset of the set of stochastic matrices corresponding to POVMs; one might interpret these restrictions as a kind of uncertainty principle. With respect to a given reference action, we call these the \emph{physically valid} regions. 

The closest $\Phi$ can approach to the identity, by any unitarily invariant norm, is when both the MIC and post-measurement states of the reference action are symmetric informationally complete POVMs (SICs) \cite{DeBrota:2018}. A SIC is a MIC with the additional properties that every effect is rank-1 and the Hilbert--Schmidt inner product between distinct effects is a constant~\cite{Zauner1999,Renes2004,Fuchs2017}. For $\mathcal{H}_d$, this implies a SIC $\{H_i\}$ satisfies 
\begin{equation}
    \tr H_iH_j=\frac{d\delta_{ij}+1}{d^2(d+1)}\;.
\end{equation}
An example of a $d=2$ SIC is given explicitly by $H_i=\frac{1}{2}\ketbra{\psi_i}{\psi_i}$ where
\begin{equation}
\ket{\psi_1}=\begin{pmatrix}
\sqrt{\frac{1}{6}\left(3+\sqrt{3}\right)}\\
\sqrt{\frac{1}{6}\left(3-\sqrt{3}\right)}e^\frac{i\pi}{4}
\end{pmatrix}\;,
\end{equation}
$\ket{\psi_2}=\sigma_z\ket{\psi_1}$, $\ket{\psi_3}=\sigma_x\ket{\psi_1}$,  $\ket{\psi_4}=\sigma_x\sigma_z\ket{\psi_1}$, and $\sigma_i$ are the Pauli matrices. These states form a regular tetrahedron in the Bloch sphere. If this SIC is the POVM for a reference action, the reference probability vector $p(i)=\tr \rho H_i$ for an arbitrary qubit 
\begin{equation}
    \rho = \frac{1}{2}\begin{pmatrix}
    1+c && a-ib\\
    a+ib && 1-c
    \end{pmatrix}
\end{equation}
where $|\!|(a,b,c)|\!|\leq 1$ is
\begin{equation}
p=\frac{1}{12}\begin{pmatrix}
3+\sqrt{3}(a+b+c)\\
3+\sqrt{3}(-a+b-c)\\
3+\sqrt{3}(a-b-c)\\
3+\sqrt{3}(-a-b+c)
\end{pmatrix}\;.
\end{equation}
The physically valid region for this reference measurement thus corresponds to the ball inscribed in the 4-outcome probability simplex. 

If $\rho_i=\ketbra{\psi_i}{\psi_i}$ are the post measurement states of a reference action, then, for example, the conditional probability matrix $R(j|i)=\tr D_j\rho_i$ for the Pauli $X$ measurement, which has effects
\begin{equation}
D_1=\frac{1}{2}\begin{pmatrix}
1&&1\\
1&&1
\end{pmatrix} \quad \text{and} \quad
D_2=\frac{1}{2}\begin{pmatrix}
1&&-1\\
-1&&1
\end{pmatrix},
\end{equation}
is given by
\begin{equation}\label{paulix}
   R_X= \frac{1}{6}
    \begin{pmatrix}
    3+\sqrt{3} & 3-\sqrt{3} & 3+\sqrt{3} & 3-\sqrt{3}\\
    3-\sqrt{3} & 3+\sqrt{3} & 3-\sqrt{3} & 3+\sqrt{3}\\
    \end{pmatrix}\;.
\end{equation}
Similarly, the Pauli $Y$ and $Z$ conditional probability matrices are
\begin{equation}\label{pauliy}
   R_Y= \frac{1}{6}
    \begin{pmatrix}
    3+\sqrt{3} & 3-\sqrt{3} & 3-\sqrt{3} & 3+\sqrt{3}\\
    3-\sqrt{3} & 3+\sqrt{3} & 3+\sqrt{3} & 3-\sqrt{3}\\
    \end{pmatrix}\;
    \end{equation}
and
\begin{equation}\label{pauliz}
    R_Z=\frac{1}{6}
    \begin{pmatrix}
    3+\sqrt{3} & 3+\sqrt{3} & 3-\sqrt{3} & 3-\sqrt{3}\\
    3-\sqrt{3} & 3-\sqrt{3} & 3+\sqrt{3} & 3+\sqrt{3}\\
    \end{pmatrix}\;.    
\end{equation}

When the MIC and post-measurement states are the \emph{same} SIC, which coincides with L\"uders rule state updating~\cite{Barnum2002,Busch2009}, we have $\Phi=(d+1)I-\frac{1}{d}J$, where $J$ is the matrix of all 1s, and equation \eqref{bornrule} becomes
\begin{equation}\label{urgleichung}
    q(j)=\sum_{i=1}^{d^2} \left((d+1)p(i)-\frac{1}{d}\right)R(j|i)\;.
\end{equation}
As the measurement and post-measurement states are SICs, the physical postulate \eqref{urgleichung} is as close to the LTP as quantum theory allows. This optimality provides some mathematical evidence for the idea, previously advanced by QBists due to this equation's aesthetic appeal, that SICs should be thought of as a ``Bureau of Standards'' reference action, ideally situated to examine what intuitions must be revised to arrive at quantum theory. 

Such pleasing simplicity does not, however, come for free --- unlike MICs in general, SICs are not yet known to exist in every finite Hilbert space dimension. There are a variety of mathematical reasons to believe they should exist (see \cite{Fuchs2017} and references), and, to our knowledge, everyone who works on the problem believes that they do. As far as explicit examples go, exact SICs are known in over 100 dimensions and high-precision numerical SICs have been found in nearly 100 more~\cite{Grassl2021}. While it would be shocking if a dimension were found where there was provably no SIC, this possibility does not bother us for two reasons. First, the number of known SICs is large enough for practical purposes in the near future. Second, as noted in \cite{DeBrota:2018b} and \cite{DeBrota2020b}, in some circumstances one may prefer to use a MIC with properties tailored to the problem at hand; for example, one with a specific tensor product structure appropriate to the experimental scenario. 

We will not dwell further on the question of choosing a particular reference action. For our purposes it is sufficient to assume a reference action has been chosen, and, in fact, in the examples we consider in the next two sections, our quantum agents use the $d=2$ SIC reference action defined above. 

We have shown that a belief in quantum mechanics and a suitable abstraction of classical mechanics may both be understood within the context of a reference action and an associated physical postulate. A rational agent who adopts the classical postulate is a \emph{classical agent} and a rational agent who adopts the quantum postulate is a \emph{quantum agent}. 

Beliefs about the physical character of the world come into play when relating expectations for different actions. A classical agent believes that a reference action reads off intrinsic, pre-existing properties of a system, while a quantum agent explicitly rejects this picture --- precisely what could compel the reference action structure a quantum agent \emph{does} believe is an open question. 

Aside from the difference in what they regard as a reference action, both types of agent behave as a rational agent does, by taking actions which maximize their expected utility. Adding a physical postulate changes nothing about this; quantizing the rational agent framework is straightforward. For a quantum agent, an action is called a quantum measurement, but like any action, an agent may prefer some consequences to others and these preferences may be assigned numerical values with a utility function. When it is clear from context, we will interchangeably use standard Hilbert space and reference action terminology for quantum agents. 

Now that we have seen how to upgrade rational agents to classical and quantum agents, in the next section we turn to the question of agents learning from experience and see how it depends upon the physical postulate.

\section{Inference}\label{sec:inference}

As we have discussed above, probabilities are guides to action. In the interest of making informed decisions, an agent may decide to conduct preliminary experiments, hoping that the probabilities conditioned on the outcomes of these tests will serve them better. Bayes rule, an important and straightforward consequence of coherence, provides guidance in this scenario.

Bayes rule states that for any mutually exclusive and exhaustive set of events $\{H_i\}$, often called \emph{hypotheses} in this context, and another event $D$, usually called \emph{data}, 
\begin{equation}\label{bayesrule}
    p(H_i|D)=\frac{p(D|H_i)p(H_i)}{\sum_j p(D|H_j)p(H_j)}\;.    
\end{equation}
The reason for the terminology comes from the use of Bayes rule in an inference scenario where it provides the coherent relation between one's \emph{prior} $p(H_i)$ for a hypothesis $H_i$ and one's \emph{posterior} $p(H_i|D)$, the probability for the hypothesis conditional on the occasion of obtaining data $D$. The other term, $p(D|H_i)$, captures the model, or \emph{likelihood}, for the data given a particular hypothesis; given $H_i$ is true, it provides one's probability of obtaining a particular data set $D$. Bayes rule thus reveals the impact of data on a rational agent's expectations.

Bayes rule makes an appearance in the intuitive setting of repeating an experiment, which is the most important case for our purposes. We will see that a simple structural assumption about an agent's joint probability distribution for measurement outcomes allows us to treat the appearance of a set of outcomes as data and correspondingly update a prior to a posterior.

Suppose an agent will take the same action on each of a sequence of systems and receive the outcomes $x_1,\ldots,x_N$. If this agent has a joint probability distribution for these outcomes, the basic identity $P(A|B)=P(A,B)/P(B)$ noted above tells us that the conditional probability for the remaining outcomes given the appearance of some subset $x_1,\ldots,x_M$ is
\begin{equation}\label{conditional}
    p(x_{M+1},\ldots,x_N|x_1,\ldots,x_M)=\frac{p(x_1,\ldots,x_M,x_{M+1},\ldots,x_N)}{p(x_1,\ldots,x_M)}\;.
\end{equation}
Having asserted a full joint distribution, what the agent learns by performing a subset of the actions is straightforward. The obvious difficulty is arriving at the full joint probability distribution $p(x_1,\ldots,x_N)$ in the first place. An agent may simply lack the capability to sort out precise expectations for a long list of outcomes, especially if they expect some outcomes will be highly correlated. However, if they regard the systems they receive to be equivalent, such that they are contemplating the outcomes of a number of repeated trials, a considerable simplification occurs.

\subsection{Exchangeability}\label{sec:infexch}

If an agent believes they will act on a sequence of operationally indistinguishable systems, they believe that any order in which they receive a given set of outcomes is equally likely, that is, any joint probability distribution they may assign for the outcomes must be permutation symmetric:
\begin{equation}
    p(x_{\pi(1)},x_{\pi(2)},\ldots, x_{\pi(N)})=p(x_1,x_2,\ldots,x_N)\;,
\end{equation}
where $\pi$ is any permutation of the set $\{1,\ldots,N\}$. Furthermore, as they could always do one more trial, the possible joint distributions are also restricted such that it is always possible to extend them to another permutation symmetric distribution with an arbitrarily larger number of outcomes. Joint probability distributions satisfying these two conditions are called \emph{exchangeable}.

The joint distribution for a set of independent and identically distributed outcomes is exchangeable, but the converse is not true. Exchangeability is a generalization we need in order to talk about learning from experience. The issue is independence. If an agent regards the outcomes of a repeated experiment to be independent, they explicitly believe the appearance of a subset to be irrelevant to the probabilities for the remaining outcomes. A distribution which is instead a mixture of identical distributions conditional on an overall parameter, that is, a joint distribution of the form
\begin{equation}\label{repthm}
    p(x_1,\ldots,x_N)=\int_{\Theta} \prod_{i=1}^N p(x_i|\pmb{\theta})dQ(\pmb{\theta})\;
\end{equation}
for $\pmb{\theta}\in\Theta$ a set of possibly vector-valued parameters, is exchangeable even though the outcomes are not independent.

In fact, the content of the celebrated de Finetti representation theorem (Corollary 1 to Proposition 4.3 in \cite{Bernardo2000}) is that an exchangeable joint distribution for real-valued $x_i$ may \emph{always} be written uniquely in the form \eqref{repthm}.\footnote{This is a generalization of de Finetti's original proof. For further extensions and a statement of the most general version, see \cite{Bernardo2000} and the references therein.} The de Finetti theorem allows an agent who regards a situation as exchangeable to act \emph{as if} there were ``unknown parameters'' $\pmb{\theta}$ for which they have a prior density $dQ(\pmb{\theta})$ and which generates the likelihood or model
\begin{equation}
    p(x_1,\ldots,x_N|\pmb{\theta})=\prod_{i=1}^N p(x_i|\pmb{\theta})\;.
\end{equation} 

In words, if an agent regards an experimental situation to be exchangeable, their joint distribution must satisfy the form \eqref{repthm}. Although not postulated, the parameters $\pmb{\theta}$ naturally appear. The practical benefit of this is that the agent likely has more immediate expectations for $\pmb{\theta}$ captured by $dQ(\pmb{\theta})$ than they do for the whole joint probability distribution. Further assumptions about the structure of measurement outcomes may then render the components of the likelihood equal to one of the familiar distributions from the statistics literature, such as the Normal distribution (with the mean and standard deviation as parameters) or the Poisson distribution (with a single parameter equal to the mean and the variance), in which case one might say the outcomes are ``sampled from a distribution''. 

Equation \eqref{repthm} simplifies in the case which will later be of most interest to us, namely when $x_i$ takes one of a finite number of values $x_i\in \{1,\ldots, k\}$. Then the parameter may be directly thought of as an ``unknown probability vector'' and the likelihood becomes a product of its components:
\begin{equation}
    p(x_1,\ldots,x_N)=\int_{\mathcal{S}_k}\prod_{i=1}^N p(x_i|\mathbf{p}) P(\mathbf{p}) d\mathbf{p}\;,
\end{equation}
where $\mathbf{p}=(p_1,\ldots,p_k)$ is a probability vector in $\mathcal{S}_k$, the $k$-outcome probability simplex, and $P(\mathbf{p})d\mathbf{p}$ is the agent's prior density over the possible ``unknown probability vectors''. 

It is now a simple matter to establish that exchangeability renders \eqref{conditional} an instance of Bayes rule where a prior belief in a parameter is related to a posterior: simply use the representation theorem on the numerator and denominator and rearrange the terms. Explicitly, the prior density $dQ(\pmb{\theta})$ becomes the posterior density (Corollary 2 to Proposition 4.3 in \cite{Bernardo2000})
\begin{equation}\label{bayesdensity}
    dQ(\pmb{\theta}|x_1,\ldots,x_M)=\frac{\prod_{i=1}^M p(x_i|\pmb{\theta})dQ(\pmb{\theta})}{\int_\Theta\prod_{i=1}^M p(x_i|\pmb{\theta})dQ(\pmb{\theta})}\;.
\end{equation}
The parameters $\pmb{\theta}$ thus play the role of the hypothesis and equation \eqref{bayesdensity} shows how a rational agent's posterior for this hypothesis conditional on measurement data is related to their prior. Compared to prior beliefs, posteriors after increasing amounts of data may become more peaked at a particular value, reflecting more precise beliefs. In other words, rational agents with an exchangeable prior may engage in \emph{parameter estimation}. When there is no cause for confusion, we may work directly with probability densities when the joint distribution is assumed to be exchangeable.

Because a probability is an expression of an agent's degree of belief, a Bayesian requires justification for something like an unknown probability or probability parameterization. It is therefore pleasing to see that the Bayes rule inferential setting emerges in an exchangeable situation without ever having to grant probabilities or parameters non-subjective status. 

\subsection{Estimating the bias of a coin}\label{sec:coin}

Consider an agent receiving and flipping coins ``prepared'' by an exogenous source. We suppose a coin is a system upon which only the action ``flip'' may be taken to obtain one of the two outcomes ``heads'' ($x=1$) or ``tails'' ($x=0$). Assuming that the agent believes the coins they receive are always prepared in the same way, the joint probability distribution for the outcomes of their flips is exchangeable. As we saw in the previous subsection, this entails their prior splits into a parameter-dependent model and a density over this parameter. For the two-outcome action of flipping a coin, the agent's exchangeable joint prior takes the form
\begin{equation}\label{bernoulliprior}
    p(x_1,\ldots,x_N)=\int_0^1 \prod_{i=1}^N \theta^{x_i}(1-\theta)^{1-x_i}P(\theta)d\theta\;.
\end{equation}
In other words, exchangeability implies the agent treats this scenario as if their outcomes are a random sample from a Bernoulli likelihood conditional on the value of a parameter $\theta$, the ``unknown probability for outcome 1 for each trial'', or, more simply, the ``bias of the coin'', for which they have a prior density $P(\theta)$. In this form, the agent's probability conditioned on the appearance of data takes the form of Bayes rule updating of the prior density,
\begin{equation}
    P(\theta|x=i)=\frac{P(x=i|\theta)P(\theta)}{\int_0^1P(x=i|\theta)P(\theta)d\theta}\;,
\end{equation}
where $P(x=1|\theta)=\theta$ and $P(x=0|\theta)=1-\theta$. Making flips thus leads to a refinement in the probability density for the bias of the coin. 
%This agent, as will all agents in our examples, has an exchangeable joint prior for the outcomes of their reference action on each system. For this first example, we suppose the agent may \emph{only} perform the reference action. 

We have not specified a physical postulate for this agent because it isn't relevant here --- the only possible action available to the agent is ``flip'', so how their beliefs for this action are related to their beliefs for other actions doesn't come into play. However, we remark that the action ``flip'' cannot be the \emph{reference} action for a quantum agent because $N=2$ is not the square of an integer (see \S\ref{sec:postulates}).

\begin{figure}[H]
\centering
\includegraphics[width=\textwidth]{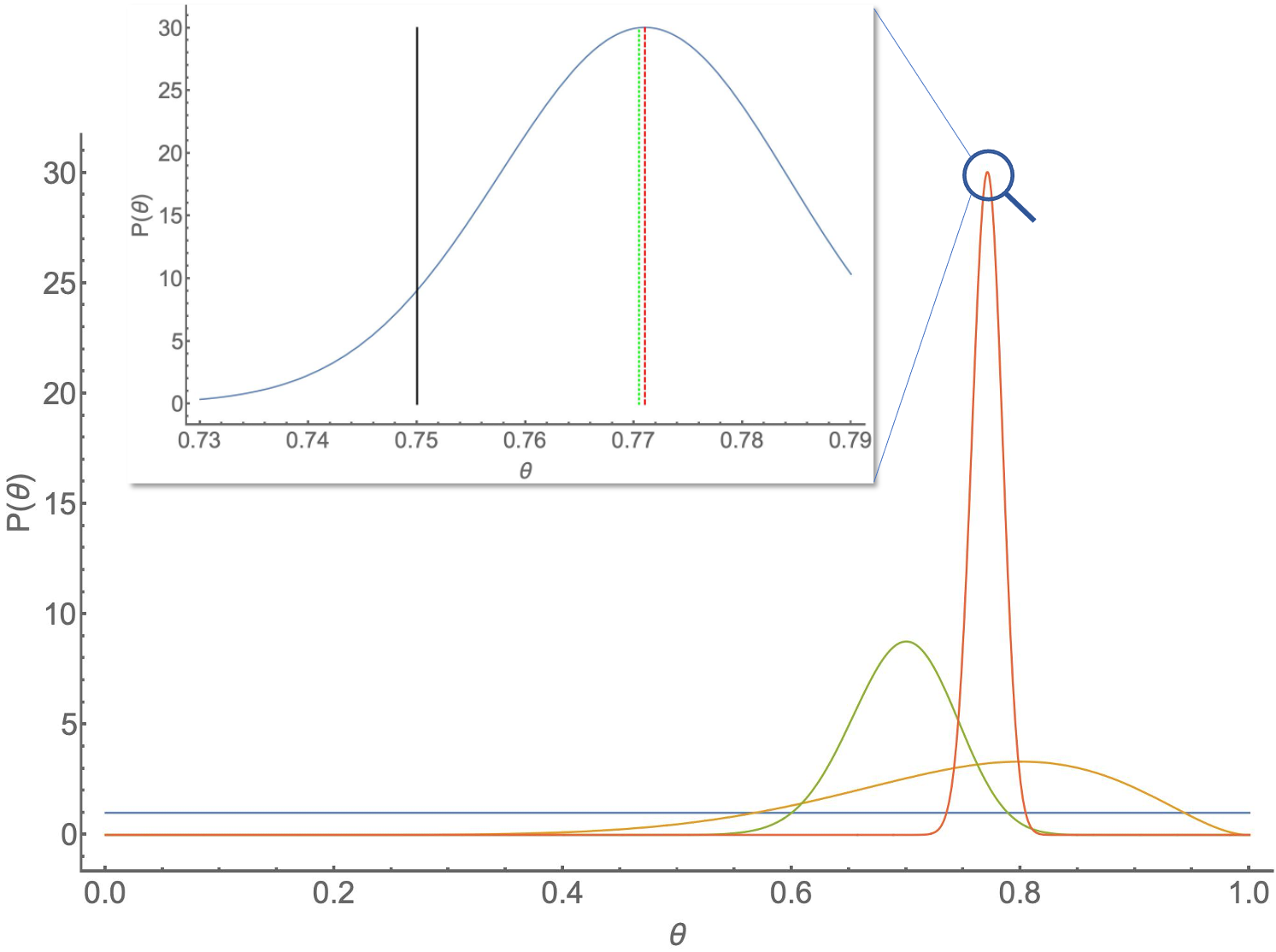}
\caption{\label{fig:bernoulliExo} Posterior densities for the parameter $\theta$, the ``bias of the coin'', after 0 (blue), 10 (orange), 100 (green), and 1000 (red) updates for a single $N=2$ classical agent flipping coins prepared by an exogenous source with bias $\theta=3/4$. The agent's initial prior is exchangeable with a uniform density for the parameter $\theta\in[0,1]$. The inset plot is a view of the posterior density after 1000 updates for a small range of $\theta$ values around the peak. The solid black line is at $\theta=3/4$, the prepared bias. The red dashed line is at $\theta=0.771$, the final proportion of heads. The green dotted line is at $\theta=0.7705$, the mean value of the agent's final posterior. As more data is collected, the posteriors become more peaked and track closely with the running frequency of heads, which, after 1000 flips, is close to the true bias of the coin.}
\end{figure}

\begin{figure}[H]
\centering
\includegraphics[width=\textwidth]{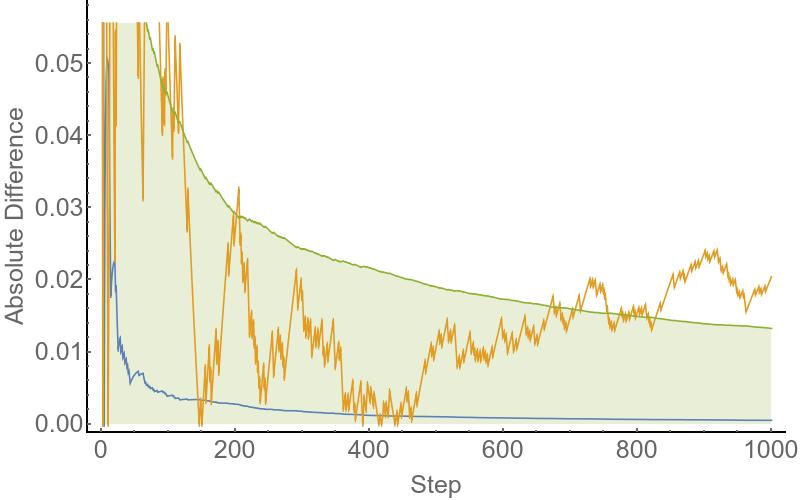}
\caption{\label{fig:bernoulliExoabsdiff} For the simulation of a single $N=2$ classical agent flipping coins prepared with $\theta=3/4$ by an exogenous source: The blue line is the absolute value of the difference between the posterior mean and the running frequency of heads. The orange line is the absolute value of the difference between the posterior mean and $\theta=3/4$. The green curve with shading below tracks the standard deviation of the agent's posterior.  As the simulation proceeds, the posterior densities become more peaked and localized near the running frequency and prepared bias.}
\end{figure}

We can simulate the process of an agent updating their beliefs about the bias of a coin by generating outcomes for them according to some set probability. Let the simulated exogenous source produce coins with $\theta=3/4$, such that the agent receives outcome 1 with probability $3/4$ and outcome 0 with probability $1/4$. Suppose the agent has a uniform initial prior for $\theta$ on the interval $[0,1]$ and that they update their current prior to the Bayes rule posterior after each flip. Figure \ref{fig:bernoulliExo} plots posterior densities for one simulation of this process after 0, 10, 100, and 1000 flips. One can see, as more outcomes are obtained, the resulting posterior density becomes more peaked. The final peak tracks closely with the data; by the end of the simulation, the agent received outcome 1 in 771 of the 1000 flips while the mean of the final posterior is $\theta=0.7705$. In Figure \ref{fig:bernoulliExoabsdiff}, the distances from the posterior mean to the running frequency and source probability are tracked and compared to the posterior standard deviation.

One could extend the above to more outcomes and different initial priors to simulate other scenarios of inference by a rational agent, but we are primarily interested in bringing reference actions and physical postulates into the fold so that we can compare learning from quantum and classical actions. In the next subsection, we will see that with an exchangeable joint reference prior agents can perform parameter estimation by performing a diversity of actions rather than a single repeated one. 

\subsection{Inference by classical and quantum agents}\label{sec:infwithpostulates}

As we saw in \S\ref{sec:refactions}, physics becomes relevant in a decision scenario when beliefs about one action are related to beliefs about another. When considering an action, classical and quantum agents strive to meet the consistency criteria that their respective physical postulates impose on the relation between reference probabilities and probabilities for other actions. What do the outcomes of actions on a subset of systems from an exogenous source imply for a classical or quantum agent's beliefs about the rest? That is, how do we incorporate physical postulates into the inference setting? 

The answer is we endow agents with a joint \emph{reference} prior. Suppose $x_i\in \{1,\ldots,N\}$ is the outcome of the reference action on system $i$. Then the joint reference prior $p(x_1,\ldots,x_M)$ is the joint probability for the outcomes of taking the reference action on each system. The agent need not actually take the reference action on any system, but, as they have adopted a physical postulate, having beliefs about the hypothetical reference action scenario will allow them learn from a sequence of actual measurements on the systems they receive. 

We first discuss this for classical agents and then for quantum agents. In each case, we first mention the completely arbitrary joint reference prior situation and then move to the special case where the agent has an \emph{exchangeable} joint reference prior. As in \S\ref{sec:infexch}, the result of the exchangeable case is that classical and quantum agents can engage in parameter estimation, where the parameter now has the physical significance of being the ``unknown probability'' or ``unknown quantum state'' prepared by the exogenous source. In other words, agents with an exchangeable joint reference prior can perform \emph{state tomography}~\cite{Blume-Kohout2010}.

\subsubsection{Inference by classical agents}

Suppose a classical agent has the joint reference prior $p(x_1,\ldots,x_M)$, $x_i\in \{1,\ldots,N\}$, for systems received from an exogenous source. What this action corresponds to depends on what they regard to be the intrinsic properties of the systems. An illustrative choice would be a discretization of a typical continuous classical phase space, as in the classical analysis of coin flipping in \cite{Diaconis2007}. In this case, the value of $x_i$ would correspond to different regions of phase space. Let $R$ be the agent's conditional probability matrix for an action $\mathcal{D}$. If the agent takes the action $\mathcal{D}$ on the first subsystem, they expect to obtain outcome $j$ with probability
\begin{equation}\label{unconditionalq}
    q(j)=\sum_{x_1=1}^N R(j|x_1)p(x_1)\;,
\end{equation}
where $p(x_1)$ is the first subsystem marginal of their joint reference prior, that is,
\begin{equation}\label{firstmarginal}
    p(x_1)=\sum_{x_2,\ldots,x_M} p(x_1,\ldots,x_M)\;.
\end{equation}
This much follows from \S\ref{sec:refactions}. The inference problem is to deduce their joint reference prior for the remaining systems conditional on the appearance of this outcome.

After taking an action on a system and obtaining an outcome, the agent will generically assign a different joint reference probability distribution. For example, if the reference action queries discrete phase space regions and the actual action were one that provided total energy information, the post-measurement reference probability distribution would reflect the agent's model of how phase space points are affected by this measurement as well as any updates consequent on correlations present in the prior. If the action is nondestructive, there is the possibility of another action on the same system, so the new joint reference probability distribution will include outcomes for the same $M$ systems. Letting $j$ be a shorthand for ``obtaining outcome $j$ of action $\mathcal{D}$ on the first subsystem'', this is a transformation 
\begin{equation}
   \mathcal{D}: p(x_1,\ldots,x_M)\xrightarrow[]{\;\; j\;\; }p_j(x_1,\ldots,x_M)\;.
\end{equation}
The solution to the inference problem, that is, the specification of the joint reference prior for the remaining systems, is then obtained by marginalizing over the reference outcomes for the system acted upon,
\begin{equation}
    p(x_2,\ldots,x_M|j)=\sum_{x_1}p_j(x_1,\ldots,x_M)\;.
\end{equation}
If the action \emph{was} destructive, the model will directly provide the inference solution as $p_j$ will already be a joint distribution over the remaining systems.

Formal solutions to the fully general inference situation as sketched above are often not practically helpful. The reason is, just as with \eqref{conditional} in \S\ref{sec:infexch}, directly asserting a joint reference probability may be prohibitively difficult. Once again, if we restrict attention to the special case of exchangeable priors, considerable simplification occurs which renders the inference scenario tractable. Conditioning on data will once again reduce to Bayes rule updating, but this time the likelihood comes from the physical postulate. 

Suppose the classical agent's joint reference probability is exchangeable. Then, by the de Finetti theorem, it has the form
\begin{equation}
    p(x_1,\ldots,x_M)=\int_{\mathcal{S}_k}\prod_{i=1}^N p(x_i|\mathbf{p}) P(\mathbf{p}) d\mathbf{p}\;.
\end{equation}
This form, where each outcome is identically distributed, albeit not independent, allows us to solve the inference problem without a specification of an explicit classical model. By the conditional probability identity, we can write
\begin{equation}\label{condprobfrac}
    p(x_2,\ldots,x_M|j)=\frac{p(j,x_2,\ldots,x_M)}{q(j)}\;,
\end{equation}
where the denominator is the unconditional probability for $j$ from equation \eqref{unconditionalq}. The first subsystem marginal, equation \eqref{firstmarginal}, for the exchangeable prior is
\begin{equation}\label{exchangeablefirstmarginal}
    p(x_1)=\int_{\mathcal{S}_k}p(x_1|\mathbf{p})P(\mathbf{p})d\mathbf{p}\;,
\end{equation}
so, plugging into \eqref{unconditionalq},
\begin{equation}\label{denominator}
    q(j)=\int_{\mathcal{S}_k}\left(\sum_{x_1}R(j|x_1)p(x_1|\mathbf{p})\right)P(\mathbf{p})d\mathbf{p}\;.
\end{equation}
Similarly, the numerator follows from an application of the physical postulate to $p(x_1|\mathbf{p})$ in the exchangeable prior:
\begin{equation}\label{numerator}
    p(j,x_2,\ldots,x_M)=\int_{\mathcal{S}_k}\left(\sum_{x_1}R(j|x_1)p(x_1|\mathbf{p)}\right)p(x_2|\mathbf{p})\cdots p(x_M|\mathbf{p})P(\mathbf{p})d\mathbf{p}\;.
\end{equation}
Thus, introducing notation reminiscent of a likelihood (the name of the term $P(D|H_i)$ in Bayes rule, equation \eqref{bayesrule}),
\begin{equation}
    p(j|\mathbf{p})\equiv \sum_{x_i=1}^N R(j|x_i)p(x_i|\mathbf{p})\;,
\end{equation}
the probability for $j$ given the reference probability vector is $\mathbf{p}$, we can divide \eqref{numerator} by \eqref{denominator} to obtain a new form for \eqref{condprobfrac}:
\begin{equation}\label{infsolution}
    p(x_2,\ldots,x_M|j)=\int_{\mathcal{S}_k}\frac{p(j|\mathbf{p})P(\mathbf{p})}{q(j)}p(x_2|\mathbf{p})\cdots p(x_M|\mathbf{p})d\mathbf{p}\;.
\end{equation}
This new form is also exchangeable with a new density for $\mathbf{p}$; in the solution to the inference problem, the prior density $P(\mathbf{p})$ is replaced by the posterior density
\begin{equation}
    P(\mathbf{p}|j)\equiv \frac{p(j|\mathbf{p})P(\mathbf{p})}{q(j)}\;,
\end{equation}
exactly as in Bayes rule. One can then iterate this process for actions on the following subsystems, using the posterior as the new prior.

\subsubsection{Inference by quantum agents}

Now we turn to the analogous quantum setting. Quantum theory prescribes specific mathematical machinery to treat the inference scenario given an arbitrary joint reference prior, that is, given a joint quantum state $\rho^{(M)}\in\mathcal{H}_d^{\otimes M}$ for $M$, $d$-dimensional, systems~\cite{Schack2001}. An arbitrary action $\mathcal{D}$ with POVM $\{D_j\}$ on some subset of these systems is specified by a choice of Kraus operators $\{A_{kl}\}$, understood to act trivially on the untouched systems, such that $\sum_lA^\dag_{kl}A_{kl}=D_k$. The joint post-measurement state after taking such an action and obtaining outcome $k$ is 
\begin{equation}
    \rho_k^{(M)}=\frac{1}{\tr \rho^{(M)}D_k}\sum_l A_{kl}\rho^{(M)}A_{kl}^\dag\;.
\end{equation}
The state of the remaining systems is then the partial trace of $\rho_k^{(M)}$ over the measured subsystems. This provides the fully general way a quantum agent's beliefs conditional on some data are related to their unconditional expectations. To cast this into the explicit form of a joint reference probability, one uses the Born rule to calculate the joint probability for a reference action with POVM $\{E_j\}$ on each subsystem. For instance, the joint reference prior is
\begin{equation}\label{rhotoprobs}
    p(x_1,\ldots,x_M)=\tr \rho^{(M)} \left(E_{x_1}\otimes\cdots\otimes E_{x_M}\right)\;.
\end{equation}

Just as for classical agents, a simplification occurs when priors are assumed to be exchangeable. Nothing about the classical analysis was specific to classical agents except for the form of the physical postulate and the associated physically valid region, in this case the whole probability simplex, where the prior density $P(\mathbf{p})$ and all subsequent posteriors live. Accordingly, the same derivation goes through for a quantum agent with the Born rule as the likelihood instead of the classical postulate equivalent to the LTP. 

The fact that exchangeability also entails a particular prior structure in the quantum setting is the content of the quantum de Finetti theorem~\cite{Hudson1976,Caves2002}. The result is that an exchangeable joint quantum state may be written uniquely in the form
\begin{equation}\label{qdf}
    \rho^{(N)}=\int_\Omega P(\rho)\rho^{\otimes N}d\rho\;,
\end{equation}
where the integration is over quantum state space $\Omega$, that is, density operators on $\mathcal{H}_d$, and $d\rho$ is a measure for that space. Just as in \eqref{rhotoprobs}, a MIC measurement on each subsystem puts this into explicit joint reference probability form, which, in this case, casts \eqref{qdf} in the form of the classical de Finetti representation; this is no coincidence, MICs were instrumental to the elementary proof in~\cite{Caves2002}. There cannot be any entanglement between the subsystems in an exchangeable prior --- this is explicitly ruled out by \eqref{qdf}. In the quantum setting, exchangeability again allows us to behave \emph{as if} each system gives outcomes by sampling a parameter, in this case the ``unknown quantum state'', for which we have a prior density $P(\rho)$. 

By analogy with the classical case, obtaining outcome $j$ for an action with POVM $\{D_j\}$ on one of the systems leads to updating the prior density $P(\rho)$ with Bayes rule. This time, however, the Born rule gives us the likelihood $p(j|\rho)\equiv \tr \rho D_j$, and the posterior density is
\begin{equation}\label{quantumbayesrule}
    P(\rho|j)=\frac{p(j|\rho)P(\rho)}{q(j)}\;,
\end{equation}
where $q(j)=\int P(\rho)p(j|\rho)d\rho$. This expression is sometimes called the ``quantum'' Bayes rule because the ``hypothesis'' is a quantum state \cite{Schack2001}.

What we have seen is that both our quantum and classical agents may learn from a sequence of arbitrary actions in an exchangeable experimental scenario. Another way to put this is that both types of agents may implement state tomography, where the goal is to assign a state, or a reference probability vector, to the systems a repeated process produces. Starting with an exchangeable prior, as either kind of agent takes actions on the systems they receive, they update their probability density for the ``unknown'' reference probability by Bayes rule with their physical postulate as a likelihood. In a pure inference scenario, forecasting the identity of a state produced by some laboratory equipment may be the end goal, and so our agents are capable of such a task. In the next subsection, we give a simulation of this scenario.

\subsection{Estimating a quantum state}\label{sec:tomography}

Here we illustrate the inference setting of the previous subsection for the case of a quantum agent. Consider a quantum agent receiving systems prepared by an exogenous source who chooses an action to take on each one. $N$ can be as small as 4, corresponding to qubit quantum systems. We suppose our agent has an exchangeable joint reference prior for the $N=4$ outcome SIC reference action introduced in \S\ref{sec:postulates}. As discussed in the previous subsection, this allows us to work with a probability density over quantum state space which is updated by Bayes rule with the Born rule likelihood upon obtaining data, equation \eqref{quantumbayesrule}. Equivalently, we may speak of a probability density over the physically valid region of the 4-outcome probability simplex, which, for a $d=2$ SIC reference measurement, is the inscribed Ball, and the physical postulate, equation \eqref{urgleichung} with $d=2$,
\begin{equation}\label{d2urgleichung}
    q(j)=\sum_{i=1}^4\left(3p(i)-\frac{1}{2}\right)R(j|i)\;.
\end{equation}

While the SIC is the reference measurement, we can imagine that it is not presently feasible for this agent to actually perform and therefore is not among their possible choices. Instead, at each step, our agent will choose between the three Pauli measurements to make. Expressed as conditional probability matrices, recall these are $R_X$, $R_Y$, and $R_Z$, equations \eqref{paulix}, \eqref{pauliy}, and \eqref{pauliz}. This is a convenient choice  because one's expectations for the Pauli measurements together fully specify a quantum state. Thus, if an exogenous source always prepares the same state and our agent chooses to make each of the Pauli measurements with sufficient frequency, we anticipate the posterior will provide a precise quantum state estimate after many iterations. This process is a Bayesian realization of quantum state tomography~\cite{Blume-Kohout2010}.

Specifically, let an exogenous source prepare systems in the state $\ket{+}=\frac{1}{\sqrt{2}}\left(\ket{0}+\ket{1}\right)$. As a reference probability vector for the SIC action, $\ket{+}$ is
\begin{equation}\label{exosource}
    \mathbf{p}=
    \frac{1}{12}\begin{pmatrix}
    3+\sqrt{3}\\
    3-\sqrt{3}\\
    3+\sqrt{3}\\
    3-\sqrt{3}
    \end{pmatrix}\;.
\end{equation}
For simplicity, we suppose the agent is indifferent to all six possible outcomes of the three actions available to them, corresponding to a uniform utility function. In this case, all three of the Pauli actions are rational, regardless of how likely the agent believes any of the outcomes are to occur: we simply choose one of the three with probability $1/3$ each time. Finally, let the agent have a uniform initial prior over all of quantum state space.

\begin{figure}[H]
\centering
\includegraphics[width=\textwidth]{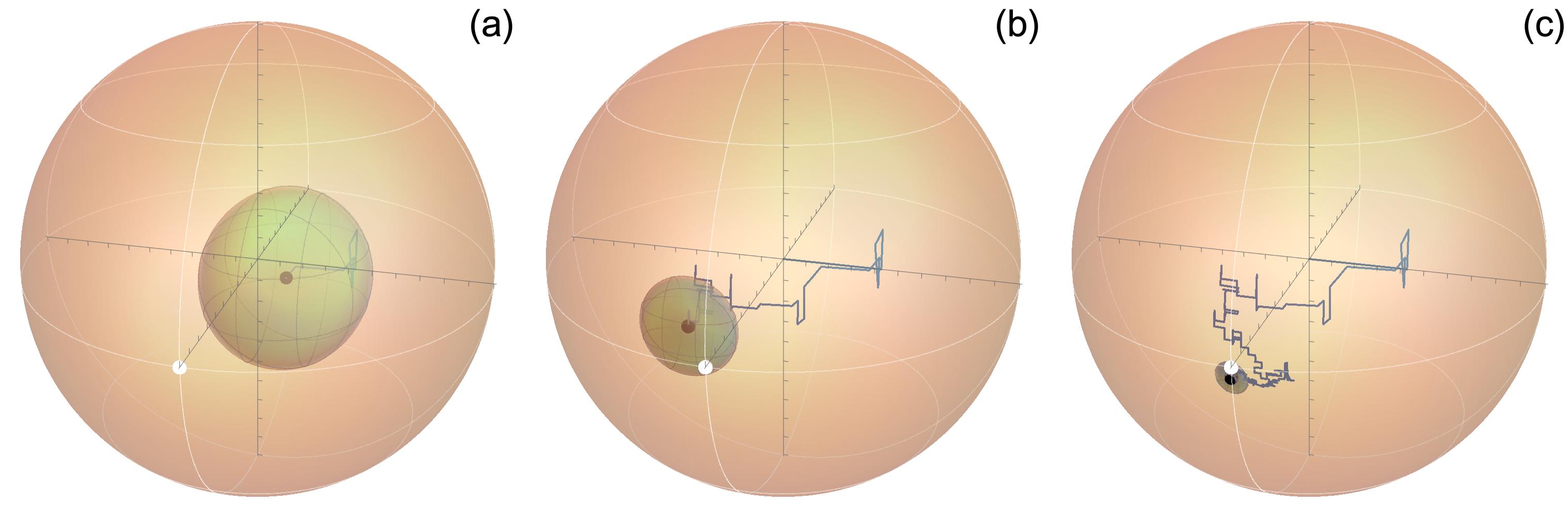}
\caption{\label{fig:qubitExo} Bloch ball plots of the posterior density within the standard deviation ellipsoid after (a) 10, (b) 50, and (c) 500 updates for a single $N=4$, i.e. qubit, quantum agent who receives systems prepared in the state $\ket{+}=\frac{1}{\sqrt{2}}\left(\ket{0}+\ket{1}\right)$ by an exogenous source and randomly takes one of the three Pauli measurements at each step. The agent's initial prior is uniform over the Bloch ball. In each plot, the black dot is the posterior mean, the blue line is the path of the previous posterior means, and the white dot is $\ket{+}$. Early on, the posterior mean is near the center of the Bloch ball, corresponding to a highly mixed state, with a large standard deviation ellipsoid. As the simulation proceeds it moves closer to the boundary, corresponding to a purer estimate, and the standard deviation ellipsoid shrinks in all directions, especially radially.}
\end{figure}

Figure \ref{fig:qubitExo} is a visualization of the agent's posteriors after 10, 50, and 500 updates. The physically valid region, quantum state space, is represented by the orange Bloch ball. A quantum agent's probability density is a distribution over this ball. For a given distribution, one can compute the covariance matrix and, from it, form the standard deviation ellipsoid whose semi-axes are the eigenvectors of the square root of the covariance matrix. Within the standard deviation ellipsoid of each posterior, we display a 3D density plot of the posterior density itself. At the center of each ellipsoid is a black dot, indicating the posterior mean. The white dot at the boundary of the ball is located at the source state $\ket{+}$. One can see the posterior after 10 updates (a) is relatively diffuse while the posterior after 500 updates (c) is sharply peaked. 

\begin{figure}[H]

\includegraphics[width=\textwidth]{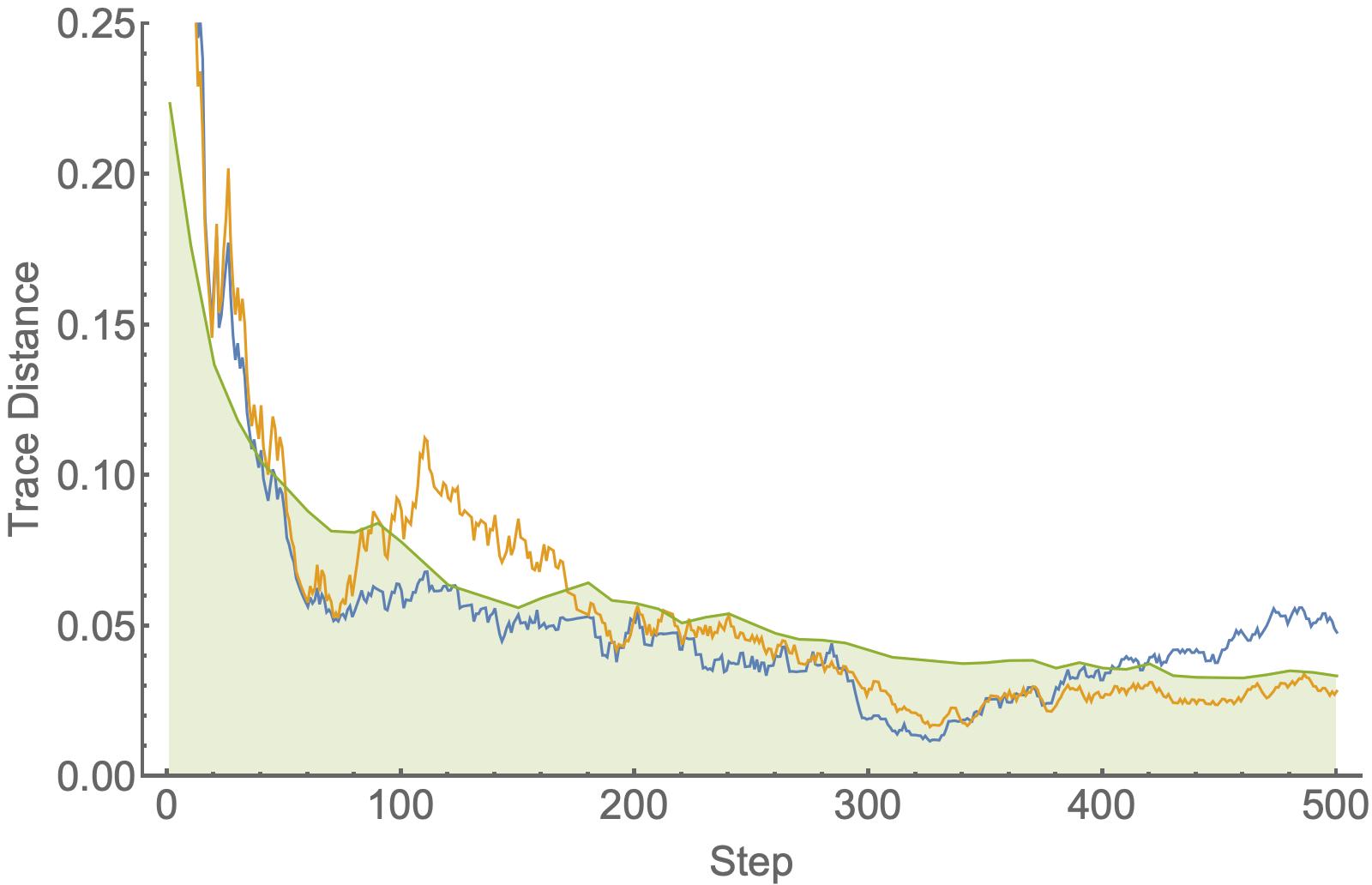}
\caption{\label{fig:qubitExoTrDist} For the simulation of a single $N=4$ quantum agent who receives qubits prepared in the state $\ket{+}$ by an exogenous source and takes a random Pauli measurement at each step: The blue curve is the trace distance between the posterior mean and the operator corresponding to the running frequency of Pauli outcomes, that is, the Hermitian, but not necessarily positive semidefinite, operator $\frac{1}{2}(I+\vec{a}\cdot\vec{\sigma})$ where the three components of $\vec{a}$ are the averages of $+1$ and $-1$ outcomes for the Pauli $X$, $Y$, and $Z$ measurements thus far taken. The orange curve is the trace distance between the posterior mean and $\ket{+}$. The green curve with shading below is the length of the semi-major axis of the standard deviation ellipsoid, calculated every 10 updates. As the simulation proceeds, the posteriors grow more peaked and localized near the running frequency and the source state.}
\end{figure}

In order to quantify convergence, we can track distances between points in operator space, such as the distance between the posterior means and $\ket{+}$. We may measure distance with one of the measures for distinguishing quantum states~\cite{Nielsen2010}. The trace distance between two quantum states $\rho$ and $\rho'$ is defined to be
\begin{equation}
    D_{\tr} (\rho,\rho')=\frac{1}{2}\tr\abs{\rho-\rho'}\;,
\end{equation}
where the absolute value of an operator is $\abs{A}\equiv \sqrt{A^\dag A}$. As shown in Figure \ref{fig:qubitExoTrDist}, the mean tracks with the data to within approximately one standard deviation of the posteriors. The distance from the mean estimate to the source state tracks similarly; in fact, while the observed frequency of data can correspond to an unphysical state, outside of the physically valid region, the source state in this example is on the boundary of state space and is actually closer to the mean estimate than the running frequency is.

One could similarly construct and simulate an example of a classical agent with an exchangeable reference prior for systems received from an exogenous source; rather than \eqref{d2urgleichung} for the physical postulate, one would use the classical physical postulate having the same form as the LTP, and rather than restricting to the Bloch ball and conditional probability matrices corresponding to POVMs, the entire probability simplex and any stochastic conditional probability matrix would be physically valid.

We will encounter this kind of agent in \S\ref{sec:interaction} where we discuss how quantum and classical agents can interact with one another. 

\section{Interacting Agents}\label{sec:interaction}
In \S\ref{sec:refactions}, we saw that rational agents may adopt physical postulates, becoming classical or quantum agents, and then, in \S\ref{sec:inference}, we saw that they can learn from experience, as exemplified by taking actions on systems received from exogenous sources of data. In this section, we finally address our strongest original motivation: interaction between agents.

What exactly should we mean by an interaction between agents of the kind we have considered? Our agents are equipped to choose between actions to take on systems presented to them. What we need, then, is for each agent to in some way affect the others by way of their mutual actions on systems. We consider the simplest setting in which only pairs of agents interact. More complex models could involve triples or larger groups of agents interacting. To simulate an interaction between two agents, we need a rule to provide each with a system and an outcome given the action they choose to take. We call such a rule an \emph{interaction}.

In our approach, it is the agents themselves, or more specifically the structure of their beliefs, which are classical or quantum. An agent's physical postulate is their personal means for relating expectations for different actions; the interaction, by contrast, is a component of the model of an agent, responsible for serving them systems and outcomes --- it is not part of their belief structure. For this reason, the choice of interaction is not necessarily related to any agent's physical postulate. What an agent expects and what actually happens to them do not \emph{need} to be coupled. This is not to say that the interaction should be unrelated to physics. The reason we believe in any scientific theory in the first place is because of sufficient agreement between expected occurrences and actual events. As long as we're mindful of this, considering interactions motivated by physical postulate strikes us as a good place to start. 

Our heuristic motivation for the kind of interactions we consider is that agents essentially broadcast their beliefs, which we treat as preparations of emitted systems, and other agents can receive these signals by taking actions on and updating their beliefs about the systems they receive. In turn, the changed beliefs of an agent will be reflected in what they send out to other agents. More specifically, we need the prior of one agent to affect the outcome of another. In the next subsection, we make this precise and propose an interaction we call expectation sampling. In the following subsections, we will simulate interactions by expectation sampling between two agents in a variety of situations.

\subsection{Expectation sampling}\label{sec:priorsampling}
For our examples we use an interaction we call expectation sampling. The idea is motivated by the intuition that interactions between agents may often be understood in terms of broadcasting and receiving signals. In a single interaction between two agents, each agent broadcasts a signal reflecting their current beliefs about the system they will receive from the other and then takes an action on the incoming system. Expectation sampling arranges for the broadcasted signal of one agent to generate an outcome for the other, given their choice of action.

Beliefs about the systems they receive are encoded in an agent's joint reference prior. To one agent, the other is essentially an exogenous source from which they receive a sequence of systems; unless independently informed, neither agent is aware of the agency of the other. For this reason, just as when an agent receives systems from a non-agential source, the case where each agent has an \emph{exchangeable} joint reference prior is an interesting and tractable place to start. We globally restrict to exchangeable joint reference priors from now on.

The simplest technical expression of an agent broadcasting a signal reflecting their current beliefs is for them to prepare and emit a system to which they assign the state corresponding to the first subsystem marginal of their joint reference probability. This is the way expectation sampling serves each agent a system. Given each agent has an exchangeable joint reference prior, the first subsystem marginal of the prior and each subsequent posterior is the mean of the parameter appearing in the mixture distribution form (See equation \eqref{exchangeablefirstmarginal}). For two classical agents exchanging coins, this means they prepare a coin they consider to be biased to their current mean estimate for the next coin they will receive. For two quantum agents exchanging two-level quantum systems, they similarly prepare qubits for the other to which they assign the state of their current mean estimate for the state of the system they will receive. 

With a system in front of them, each agent chooses an action to take as usual, by maximizing their expected utility. In doing so, they expect outcomes with probabilities derived from their physical postulate and their current estimate for the parameter value. When they actually take an action, expectation sampling generates an outcome for them with probability equal to what their physical postulate \emph{would} assign if their reference probability were the state the other agent broadcasted. 

Explicitly, consider two agents $A$ and $B$ with physical postulates $\mathcal{F}_A$ and $\mathcal{F}_B$. Let $\langle\theta_A\rangle$ and $\langle\theta_B\rangle$ denote the mean values of the parameters appearing in their current joint reference probability for the systems received from the other. Suppose $A$ takes action $\mathcal{A}$ for which the $j$th row of their conditional probability matrix is $a_j$ and that $B$ takes the action $\mathcal{B}$ with $j$th row $b_j$. When compatible in dimensionality and physically valid regions, expectation sampling generates outcome $j$ for $A$ with probability $\mathcal{F}_A(\langle\theta_B\rangle,a_j)$ and outcome $j$ for $B$ with probability $\mathcal{F}_B(\langle\theta_A\rangle,b_j)$. 

If the agents have reference actions with a different number of outcomes or their physical postulates are otherwise such that the mean of one's parameter is outside the physically valid region of the other, we can adopt a case-dependent regularization, for example a projection, to ensure the interaction is well-defined. 

With this outcome scheme in place, an interaction by expectation sampling between two agents who exchange and flip coins gives each heads with probability equal to the other's expected bias. An interaction by expectation sampling between two quantum agents, where each takes an action on a qubit quantum system received from the other, gives each agent an outcome distributed by the Born rule probabilities for their measurement on the state the other expects to receive. We will simulate these cases and more involved situations in the next subsections.

There are at least two significant advantages of the expectation sampling choice. First, it contains as a limiting case what we saw in the previous section: interactions between an agent and a non-agent. An exogenous source which ``prepares the same state'' every time may be thought of as an infinitely confident and particularly dull agent whose prior is characterized by a delta distribution at a particular parameter value. Expectation sampling is thus a natural relaxation of the single agent setting from which we might begin to identify genuine multi-agent properties.

The second advantage concerns a potential multi-agent property one may wish to investigate: agreement between agents. Some humans like to agree. Indeed, the possibility of coming to agree with someone by interacting with them can be a strong motivation for interacting in the first place. In science, it is more than merely pleasant to agree; it is often a practical necessity. Our agents are quite unlike humans, having been minimally equipped to be users of probability theory, and, through it, of physical theory. If our agents aspire to something like science, it is important to understand when, if ever, agreement between them can be reached. 

What does agreement mean for our type of agent? We focus on a kind of agreement we call probabilistic agreement, characterized by two agents coming to have similar beliefs about the likelihood of events given potential actions. After sufficient interaction, two agents probabilistically agree if their reference probability distributions have converged in some manner. As the only thing that can update in a expectation sampling interaction is the reference probability distribution, this notion of agreement seems appropriate. Going forward, all agreement is probabilistic agreement.

So does interaction by expectation sampling between two agents lead to belief convergence? Note that these beliefs are about systems received from the other, rather than from a third, static, exogenous source. Each agent's posterior will track with the data they receive, but there is no ``true'' state to compare with as there was when receiving systems from an exogenous source. There is no ``matter of fact''; agents are exchanging data on beliefs. Therefore, even from identical initial priors, it might be possible that interactions typically cause two agents' probability distributions to diverge. The second advantage of expectation sampling is that it seems to be well-suited to avoiding this possibility; as we shall see, at least some of the time, we see agents coming to agreement under this interaction.

In the next subsections, we conduct numerical investigations of interactions by expectation sampling, showcasing several variations, such as agents possessing different priors, different utilities, and different physical postulates. We identify three broad questions to help organize analysis for the situations we consider and to guide future research:
\begin{enumerate}
    \item[Q1:] Does interaction by expectation sampling cause the individual beliefs of two agents to converge to a steady state?
    \item[Q2:] Do these beliefs indicate agreement?
    \item[Q3:] What is the distribution of resulting beliefs?
\end{enumerate}
These questions remain largely open, even in the simplest cases.

The organization of the remaining subsections is as follows. In \S\ref{sec:bernoulli}, we simulate the simplest case of an interaction by expectation sampling: two classical agents who exchange and flip coins. In this case, we are additionally able to analytically demonstrate that some characteristics of agreement should be expected. In \S\ref{sec:2qubit}, we consider the simplest pair of two quantum agents interacting by expectation sampling, where we specifically fix their reference actions to be the $d=2$ SIC introduced in \S\ref{sec:postulates}. These agents choose to make one of the three Pauli measurements. We simulate two cases, one where both agents have flat utility functions and correspondingly take random actions and another where one agent has nontrivial preferences among the possible outcomes. Finally, in \S\ref{sec:qc}, we consider two distinct types of interaction by expectation sampling between a quantum and a classical agent: one where the classical agent is conceptually ``classical'' because they lack sufficient experimental capabilities and another where they are classical because they simply disbelieve the quantum physical postulate. For the latter kind of interaction, we simulate two scenarios which differ in the actions available to the classical agent. For the first simulation, the conditional probability matrices for the classical agent's actions are the same as the ones the quantum agent assigns for the Pauli measurements. For the second simulation, they are instead measurements related to the Paulis, but which cannot correspond to any actual quantum action. 

\subsection{Exchanging and flipping coins}\label{sec:bernoulli}

As mentioned above, the simplest example of interaction by expectation sampling is two agents exchanging and flipping coins, each time sending a coin with the bias they have previously estimated and updating their prior for this expectation after each flip. Here we examine this scenario analytically and then simulate two instances of it.

Suppose two classical agents, Alice and Bob, are exchanging and flipping coins. Recall from \S\ref{sec:coin} that a coin is a system upon which only the action ``flip'' may be taken to obtain outcomes ``heads'' ($x=1$) or ``tails'' ($x=0$). Although we call them classical, here Alice and Bob are most appropriately thought of as merely rational agents, rather than quantum or classical, because our notion of a coin can only be acted upon in one way. Recall that the parameter $\theta$ appearing in an exchangeable joint reference probability can be interpreted as the bias of a coin, that is, the probability for heads. As their joint probabilities are exchangeable, Alice and Bob's beliefs are captured by their probability densities for this parameter, $P_A(\theta)$ and $P_B(\theta)$.

%Now that we've seen illustrations of an agent receiving data from an exogenous source, we consider two interacting agents. Suppose there are two classical agents, Alice and Bob, sending and receiving systems to and from one another for which they each have an $N=2$ outcome reference action. Let $x\in \{0,1\}$ label the outcomes of each of their reference actions. As the single agent did in the previous subsection, when contemplating an action on a single system, these agents regard their probabilities $p(x=1)$ and $p(x=0)=1-p(x=1)$ to fully fix their expectations. For a first illustration of two agents interacting, we again remove the element of choice; Alice and Bob may \emph{only} perform the reference action. 

%We still suppose Alice and Bob each have exchangeable joint priors for the sequence of reference actions they take on systems received from the other. The fact we're considering two agents is only explicit to us; Alice, for example, may have no idea that Bob is another agent. 

%As before, exchangeability ensures both joint priors take the form of \eqref{bernoulliprior}. We call this type of agent a Bernoulli agent because they believe outcomes of their reference action are random samples from a Bernoulli likelihood. Conditioned on data, each agent updates a prior density for the parameter $\theta$ to a posterior density with Bayes rule. Subscripts $A$ and $B$ denote Alice and Bob's probabilities and probabilities densities.

If Alice receives heads, she updates her prior density by Bayes rule to the posterior density $P_A(\theta|x=1)$, as in \S\ref{sec:coin}, and similarly for tails. According to expectation sampling, she receives heads from a flip with probability equal to the mean estimate of Bob's density, $\langle \theta_B\rangle$. Thus, the expected density for Alice's posterior following the interaction is 
\begin{equation}\label{exppos}
\begin{split}
    \text{ExpPos}_A(\theta)&\equiv p_B(x_1=1) P_A(\theta|x_1=1)+ p_B(x_1=0)P_A(\theta|x_1=0)\\
    &=\left(\frac{\langle \theta\rangle_B}{\langle \theta\rangle_A}\theta+\frac{1-\langle \theta\rangle_B}{1-\langle \theta\rangle_A}(1-\theta)\right)P_A(\theta)\;,
    \end{split}
\end{equation}
and the expected density for Bob's posterior is the same with $A\leftrightarrow B$. 

One way to assess how agents' beliefs change is to examine how the mean parameter value of their posterior densities change. In this case, this would be an analysis of how each agent's mean estimate for the bias of the coins sent by the other changes with interaction. Before the interaction, Alice and Bob's bias estimates will generically differ by some amount. Let $\langle\text{ExpPos}_A(\theta)\rangle$ denote Alice's expected posterior bias estimate. From equation \eqref{exppos}, it is straightforward to compute
\begin{equation}
    \langle\text{ExpPos}_A(\theta)\rangle=\frac{(\langle\theta\rangle_B-\langle\theta\rangle_A)\langle\theta^2\rangle_A+(1-\langle\theta\rangle_B)\langle\theta\rangle_A^2}{(1-\langle\theta\rangle_A)\langle\theta\rangle_A}\;,
\end{equation}
and then
\begin{equation}
\abs{\langle\theta\rangle_B-\langle\theta\rangle_A}\geq \abs{\langle\theta\rangle_B-\langle\text{ExpPos}_A(\theta)\rangle}
\end{equation}
follows from $\langle\theta\rangle^2_A\leq\langle\theta^2\rangle_A$. The equivalent result also holds for $A\leftrightarrow B$. In words, for each agent, the expected posterior bias estimate is closer or equidistant to the other agent's prior bias estimate as a result of interaction. Iterating this process with generic overlapping initial priors, the agents should eventually come to have close posteriors after many interactions, signifying agreement. If their initial priors are disjoint, one expects many interactions to eventually lead to posteriors as close as the prior support allows. 

\begin{figure}[H]
\centering
\includegraphics[width=1.0\textwidth]{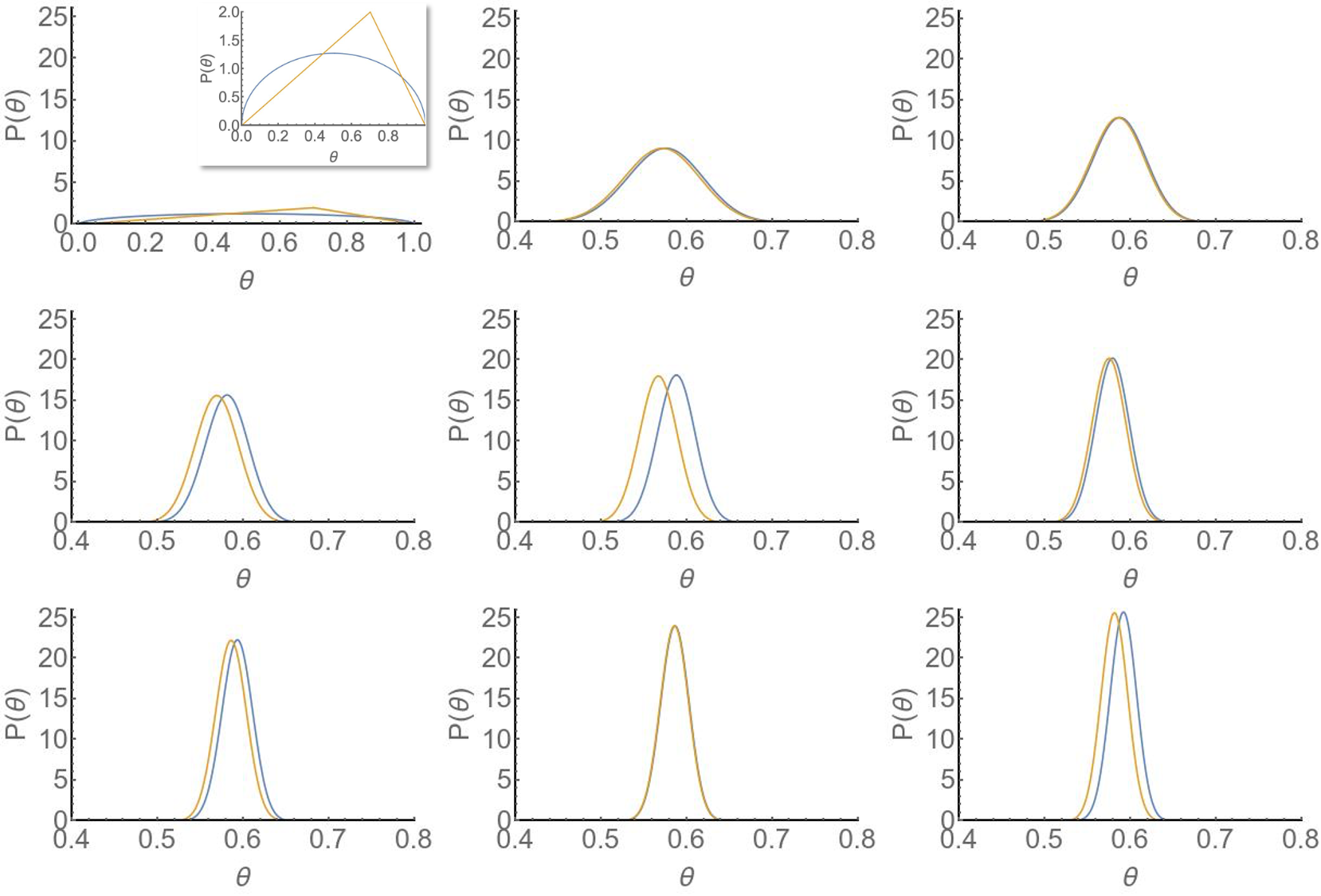}
\caption{\label{fig:wsctri} From left to right and top to bottom: Posterior densities for the parameter $\theta$, the ``bias of the coin'', for two $N=2$ classical agents interacting by expectation sampling who flip each coin they receive from the other and who have different initial priors, one with the Wigner semicircular distribution and the other with a triangular distribution peaked at $\theta=0.7$, in steps of 125 interactions, from 0 interactions in the top left to 1000 interactions in the bottom right. The subfigure plot after 0 interactions displays the initial prior plotted with a smaller range so that the features are visible. In all but the first plot, the domain is restricted to $\theta\in[0.4,0.8]$. Note that the posteriors grow more peaked with more interactions and tend to approach one another, although not necessarily monotonically, for instance, the separation in the center plot after 500 interactions.}
\end{figure}

\begin{figure}[H]
\centering
\includegraphics[width=\textwidth]{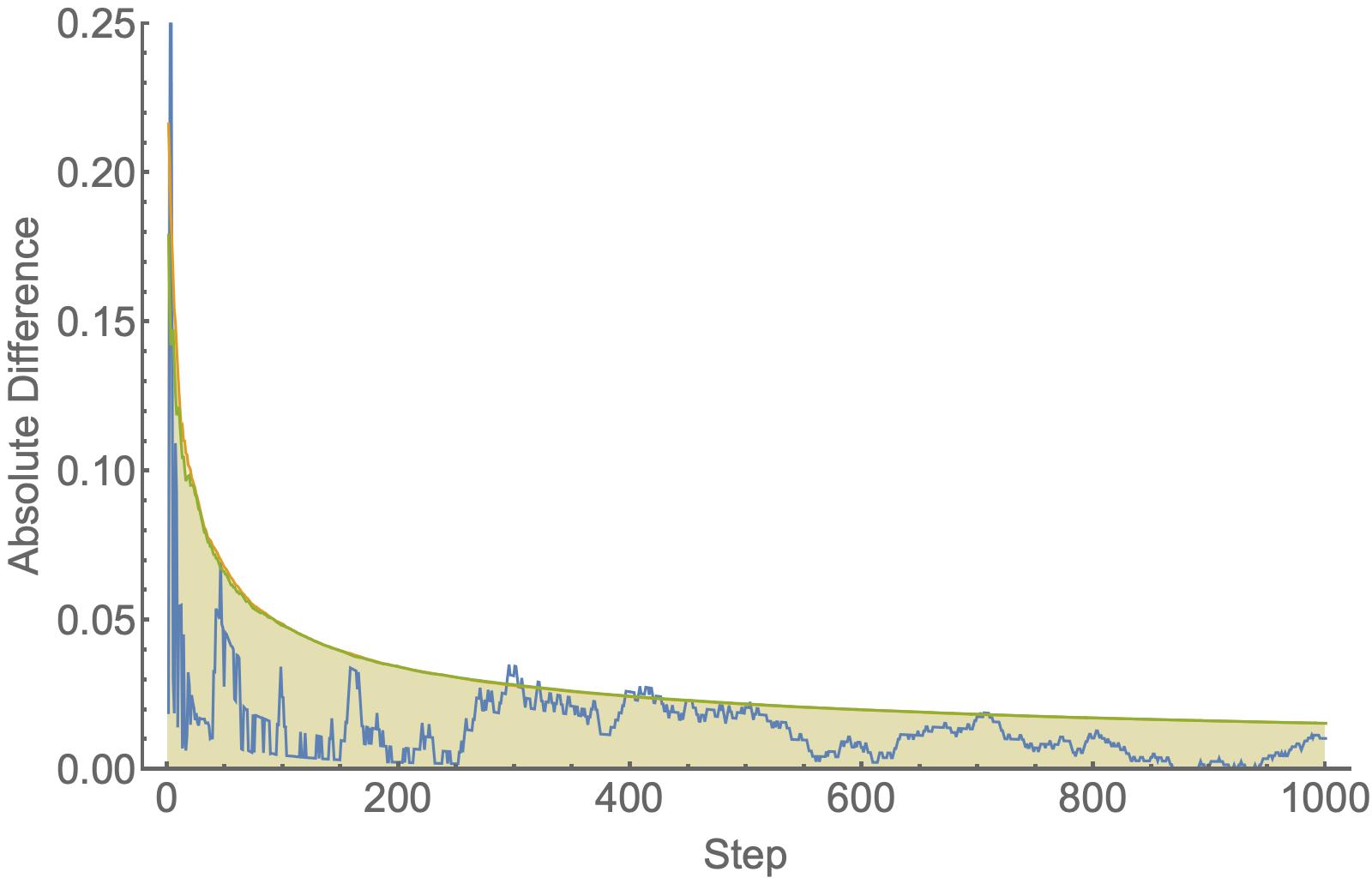}
\caption{\label{fig:wsctri-meandist} For the simulation of two $N=2$ classical agents interacting by expectation sampling who have different initial priors, one with the Wigner semicircular distribution and the other with a triangular distribution: The blue curve is the distance between the mean values of the posterior densities of each agent. Also shown are two nearly overlapping orange and green curves with shading below. The orange curve is the standard deviation for the agent with the Wigner semicircle initial prior and the green curve is the standard deviation for the agent with the triangular initial prior. The trend toward agreement is clear; there are oscillations in the mean difference but these are largely confined within the standard deviation envelope.}
\end{figure}

Figure \ref{fig:wsctri} displays this first intuition with a simulation of 1000 interactions between Alice and Bob with two quite different initial prior densities: the triangular distribution and the Wigner semicircular distribution. Figure \ref{fig:wsctri-meandist} plots the absolute value of the distance between their means alongside the standard deviations of the posteriors for each interaction. 

The simulation suggests affirmative answers to Q1 and Q2 as the final posteriors of each agent seem to be concentrating at a stationary point and the trend towards their agreement is clear. Q3 in this case will be the subject of future research; from other runs of the same simulation, we find the parameter location of this agreement can vary widely, but do not yet have a sense of what kind of distribution of agreement should be expected. 

\begin{figure}[H]
\centering
\includegraphics[width=\textwidth]{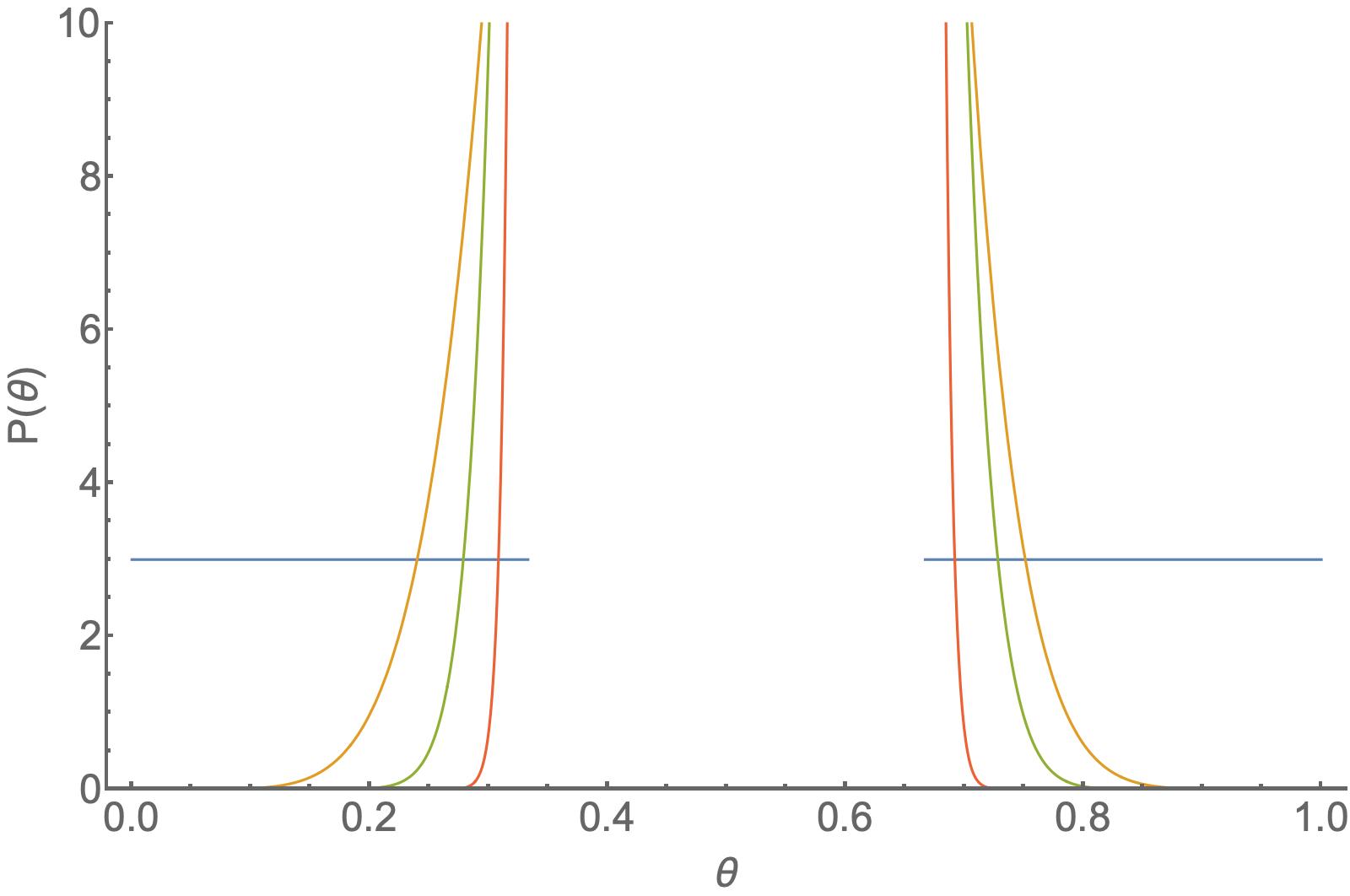}
\caption{Plot of posterior densities for the parameter $\theta$ of two $N=2$ classical agents interacting by expectation sampling who initially have uniform priors on the disjoint intervals $\theta\in[0,1/3]$ and $\theta\in[2/3,1]$ after 0 (blue), 10 (orange), 30 (green), and 100 (red) interactions. Bayes updating cannot change the support of the initial priors, but as the number of interactions increases, the posteriors become more peaked and grow closer within the bounds of the constraint on the supports.}
\label{fig:disjoint}
\end{figure}

Figure \ref{fig:disjoint} displays the latter intuition, where initial priors are disjoint, for 100 interactions. In this case, Alice and Bob have uniform prior densities over the disjoint intervals $[0,1/3]$ and $[2/3,1]$, respectively. As the number of interactions increases, their posterior densities crowd up against the boundary of the initial support; agreement is fundamentally constrained by initial priors of this sort.

A stronger characterization of agreement comes from a statistical distance measure between the expected posterior densities for the agents. Simulations suggest that convergence of this type for a range of priors and distance measures should be expected. If Alice and Bob both have \emph{uniform} initial prior densities and interact an arbitrary number of times, it is proven in Appendix \ref{sec:proof} that their expected posterior densities for a further interaction are closer than or equal to each other than their prior densities were as measured by the Kolmogorov distance measure. This result says that if Alice and Bob are both initially maximally uncertain about the bias of coins received from the other, one expects their distributions will genuinely converge in a large interaction limit. We suspect this result can be further strengthened with a relaxation of the initial condition.

\subsection{Qubit agents}\label{sec:2qubit}
Next we consider two interacting quantum agents. We will now allow our agents to take more than one action. Being able to take more than one is crucial for our agents to be ``decision makers''; if we demand they take the same action every time, there is no choice involved.

Let Alice and Bob be quantum agents with the SIC reference action interacting by expectation sampling. Each time they receive a system, they both choose one of the three Pauli measurements and have uniform utility functions across all six outcomes. As they equally value all of the possible outcomes, all choices are rational, irrespective of their expectations, so we randomize their choice: for each interaction they take one of the Pauli actions with probability $1/3$. Alice and Bob each initially have uniform priors over the Bloch ball.

\begin{figure}[H]
\centering
\includegraphics[width=\textwidth]{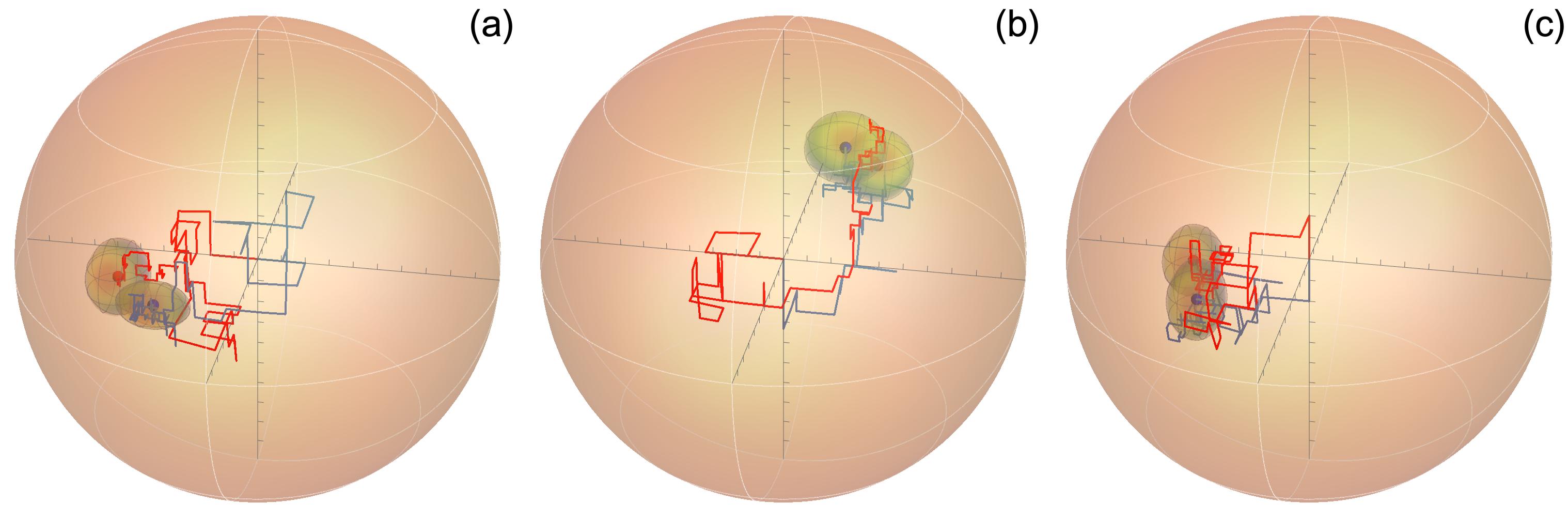}
\caption{\label{fig:qpflatU} Bloch ball plots of the final posterior densities within the standard deviation ellipsoids for three simulations, (a), (b), and (c), of 100 interactions by expectation sampling between two $N=4$, i.e. qubit, quantum agents who, for each interaction, take one of the three Pauli measurements on the system received from the other. Both agents have uniform utility functions across all six outcomes of these three actions; hence, at each interaction, both choose a Pauli measurement randomly with equal probability. Each agents' initial prior is uniform over the Bloch ball. In each plot, the blue and red dots are the means of the final posteriors and the blue and red lines are the paths of the previous posterior means. By the end of each simulation, the final means and standard deviations are similar, that is, it seems the agents are coming to agreement. Where their final posteriors end up depends on their choices and their outcomes.}
\end{figure}

\begin{figure}[H]
\centering
\includegraphics[width=1.0\textwidth]{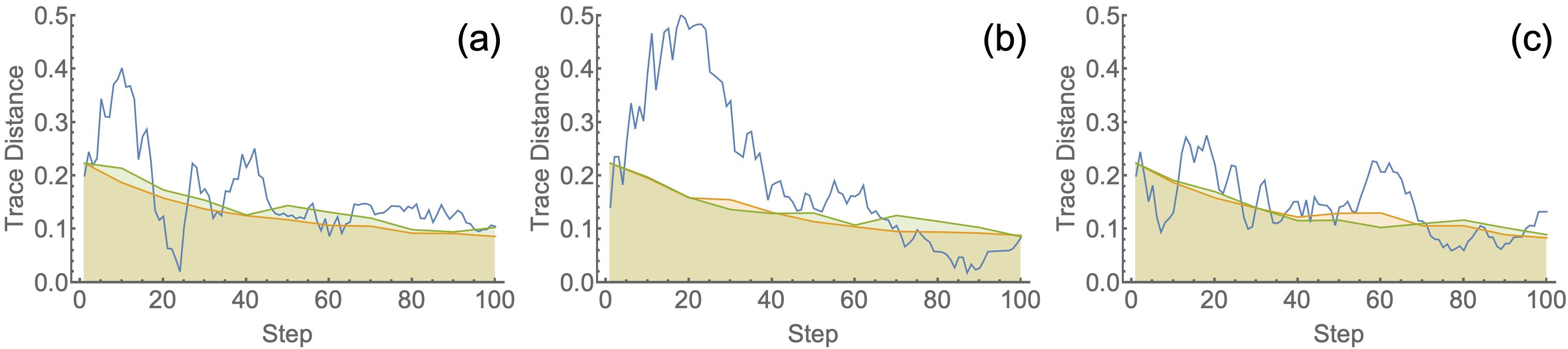}
\caption{\label{fig:qpflatU-trdist} For the simulations (a), (b), and (c) of 100 interactions by expectation sampling between two $N=4$ quantum agents who take Pauli measurements and have uniform utility functions and initial priors: The blue curve is the trace distance between the posterior means of the two agents. The green and orange curves are the lengths of the semi-major axes of the standard deviation ellipsoids, computed every 10 interactions. In each simulation there is a clear downward trend in the distance between the means and in the widest axis of the standard deviation ellipsoid.}
\end{figure}

Density plots in the Bloch ball of the final posteriors within each agents' standard deviation ellipsoid for three simulations of 100 interactions is shown in Figure \ref{fig:qpflatU}. As in \S\ref{sec:coin}, we note that the Pauli actions form an informationally complete set, so with enough outcomes from each action, we expect the agents' posteriors to be sharply peaked at some parameter value. In each case, after 100 interactions, the final posteriors for each agent are indeed highly concentrated, as the standard deviation ellipsoids are small. Additionally, the final posteriors for each agent correspond to similar regions of the valid states. In Figure \ref{fig:qpflatU-trdist}, we plot the trace distance between the means of agents' probability densities after each interaction and compare this distance to the semi-major axis of each standard deviation ellipsoid. 

The posteriors of each agent are certainly growing more peaked and closer together with more interactions. As one grows peaked, the chance the other posterior will shift significantly decreases as the most probable outcomes will correspond to a running frequency near the mean of the first. This suggests affirmative answers to Q1 and Q2. From the three simulations run, one sees the locus of agreement may vary. Q3 asks what the distribution of agreements will look like; addressing this numerically would require many more simulations. There may be special cases which can be analytically treated as well. 

Now suppose that Bob has a nonuniform utility function over the set of six Pauli outcomes. Consequently, he no longer makes purely random measurements, but instead chooses which Pauli measurement to make by picking one that maximizes his expected utility, characteristic of a rational choice (see \S\ref{sec:rationalagents}). Specifically, suppose Bob's utility function is such that the positive eigenvalue outcome of the Pauli Z measurement has utility value 0.98, the negative eigenvalue has utility 1.02, and the eigenvalues of X and Z all have utility 1. Alice remains indifferent among all six outcomes. As he updates his probability density, he may prefer to take the Z action or prefer not to take it, depending on his previous outcomes. As before, in all cases where there is a tie in the maximum expected utility, our agents make a random choice. 

\begin{figure}[H]
\centering
\includegraphics[width=\textwidth]{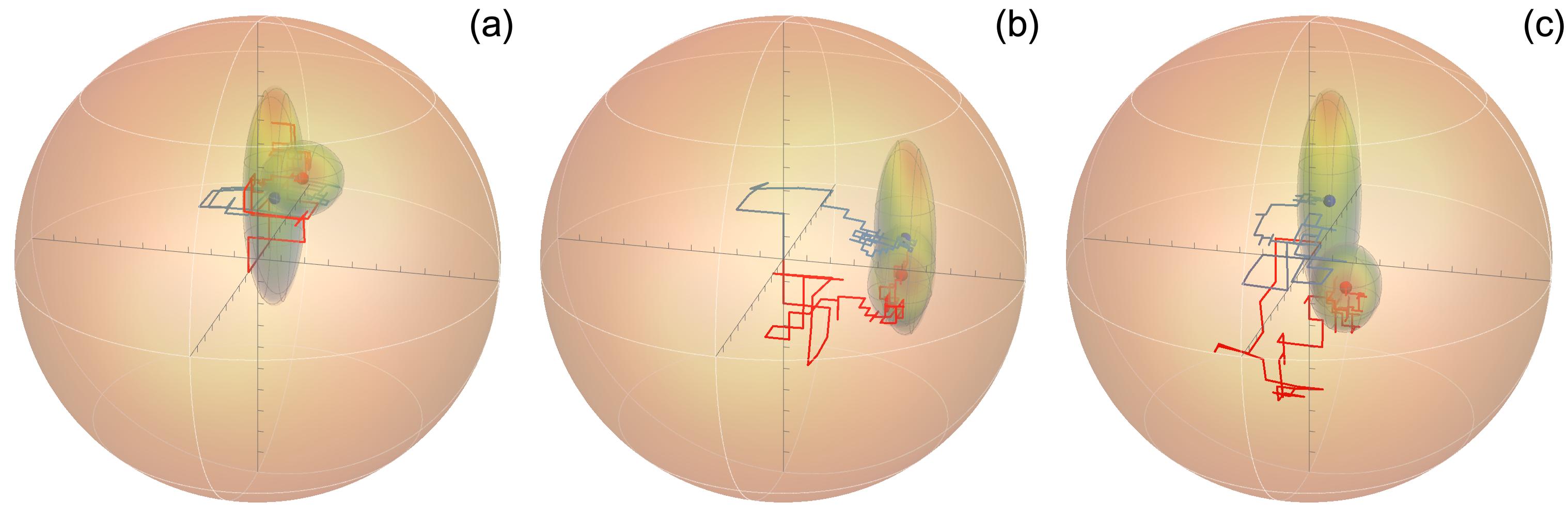}
\caption{\label{fig:qpdiffU} Bloch ball plots of the final posterior densities within the standard deviation ellipsoids for three simulations, (a), (b), and (c), of 100 interactions by expectation sampling between two $N=4$ quantum agents who, for each interaction, take one of the three Pauli measurements on the system received from the other and have \emph{different} utility functions. One agent, Alice, has a uniform utility function across all six outcomes and thus always chooses a Pauli measurement randomly with equal probability. The other agent, Bob, has utility $1$ for each of the four outcomes of the Pauli $X$ and $Y$ measurements and utilities $0.98$ and $1.02$ for the positive and negative eigenvalue outcomes for the Pauli $Z$ measurement, respectively. Each agents' initial prior is uniform over the Bloch ball. In each plot, the red dot is the mean of Alice's final posterior and the blue dot is the mean of Bob's final posterior. The red and blue lines track their previous posterior means. Alice ends up with tightly peaked final posteriors while Bob's final posteriors are smeared out due to preferring not to make the $Z$ measurement, based on the posteriors obtained from the first few interactions. They appear to be approaching agreement in the $x$ and $y$ directions.}
\end{figure}

\begin{figure}[H]
\includegraphics[width=\textwidth]{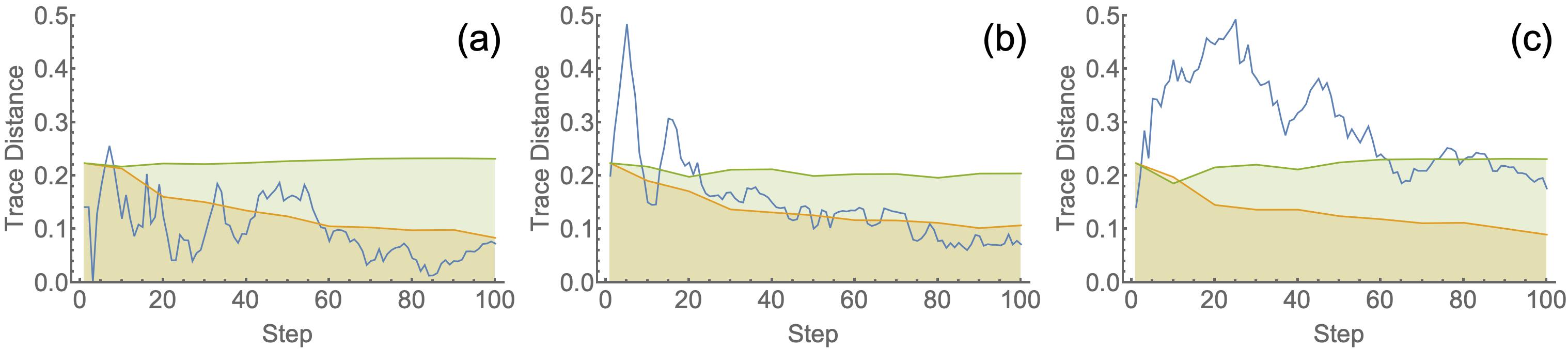}
\caption{\label{fig:qpdiffU-trdist} For the simulations (a), (b), and (c) of 100 interactions by expectation sampling between two $N=4$ quantum agents who take Pauli measurements and have uniform initial priors, but have different utility functions: The blue curve is the trace distance between the posterior means of the two agents. The orange curve is the length of the semi-major axis of the standard deviation ellipsoid for Alice, the agent with a uniform utility function, computed every 10 interactions. The green curve is the length of the semi-major axis of the standard deviation ellipsoid for Bob, the agent with a nonuniform utility function, computed every 10 interactions. Bob's semi-major axis remains more or less flat across the simulation because he avoids making the Z measurement early on. Despite this, there still appears to be a trend towards overall agreement as measured by the distance to the means, although this is a smaller effect than when both agents have flat preferences as in Figures \ref{fig:qpflatU} and \ref{fig:qpflatU-trdist}.}
\end{figure}

For this variation, density plots in the Bloch ball of the final posteriors within the standard deviation ellipsoids for three simulations of 100 interactions are shown in Figure \ref{fig:qpdiffU}. Note that Bob's posteriors display uneven narrowing as compared to Alice's, but in the directions narrowing does occur, it generally tracks with Alice's distribution. Figure \ref{fig:qpdiffU-trdist} plots the trace distance between the means after each interaction compared to the semi-major axis of each standard deviation ellipsoid. There is a trend toward agreement in the $x$ and $y$ directions, but it does not appear that we can expect $z$ agreement. In each simulation, after a few interactions, Bob began to avoid making the $z$ measurement. Without making this measurement, his posterior cannot narrow further in that direction.

\subsection{Interacting quantum and classical agents}\label{sec:qc}

The goal of our final examples is to show that it is possible to arrange for two agents with different physical postulates to interact by expectation sampling. Specifically, we consider cases of interactions between quantum and classical agents. 

As we have seen, being a classical or a quantum agent is a matter of the physical postulate adopted; a classical agent's physical postulate necessarily takes the form of the LTP, while a quantum agent's physical postulate is the Born rule, which takes the form \eqref{bornrule} where $\Phi$ is fixed by the reference action. Recall that, as a result of this, if the agent is classical, any $N$-outcome reference probability distribution is physically valid, while if they are quantum, quantum state space corresponds to a proper subset of the $N$-outcome probability simplex. How do we arrange for a quantum agent and a classical agent to interact?

Right away we see a potential issue hinted at in \S\ref{sec:priorsampling}: What if the mean value of one agent's probability density is outside the physically valid region of the other? If not regularized in some way, the normal procedure for producing outcome probabilities could result in improper probabilities. We have to take extra care in defining an interaction between different species of agent.

There are at least two distinct and interesting senses in which a quantum and a classical agent could interact, corresponding to the cardinality of the classical agent's reference action. The quantum agent's reference action will have $N=d^2$ outcomes for some integer $d\geq 2$. We interpret an interaction between a quantum agent and a classical agent with either $N=d$ or $N=d^2$ as follows. 

Suppose Quinn is a quantum agent and Clark is a classical agent with an $N=d$ outcome reference action. Clark judges the degrees of freedom of the systems he receives to be a square factor fewer than Quinn does for the systems she receives. This could conceptually mean Clark hasn't discovered quantum mechanics, possibly due to lacking sufficient experimental capabilities. Perhaps he notices that with the appropriate preparations and measurements, he can predict certain outcomes with probability one; after finding $d$ such possible configurations, he postulates that this maximal set of perfectly distinguishable states corresponds to elements of reality which his measurement reveals. 

To enable an interaction by expectation sampling between Quinn and Clark when his beliefs correspond to a classical orthonormal subtheory as above, we need to choose how the parameter estimate of one can provide a system compatible with the other's physical postulate. Suppose $d=2$. Quinn broadcasts qubit systems prepared in the state of her current mean parameter value. In line with the intuition of the previous paragraph, for the purposes of generating an outcome, Clark's reference action could correspond to performing the Pauli Z measurement. Geometrically, this regularization is a projection of Quinn's physically valid region onto Clark's region. In the other direction, we choose the state broadcasted by Clark to be $\langle\theta_C\rangle\ketbra{0}{0}+(1-\langle\theta_C\rangle)\ketbra{1}{1}$, where $\theta_C$ is the parameter in Clark's probability density. In this way, he broadcasts a state on the $z$ axis of the Bloch ball corresponding to his current mean estimate for the outcome of his two-outcome reference measurement. This arrangement is the same as the one between Q and Captain Picard in \cite{Meyer1999}.

\begin{figure}[H]
\centering
\includegraphics[width=1.0\textwidth]{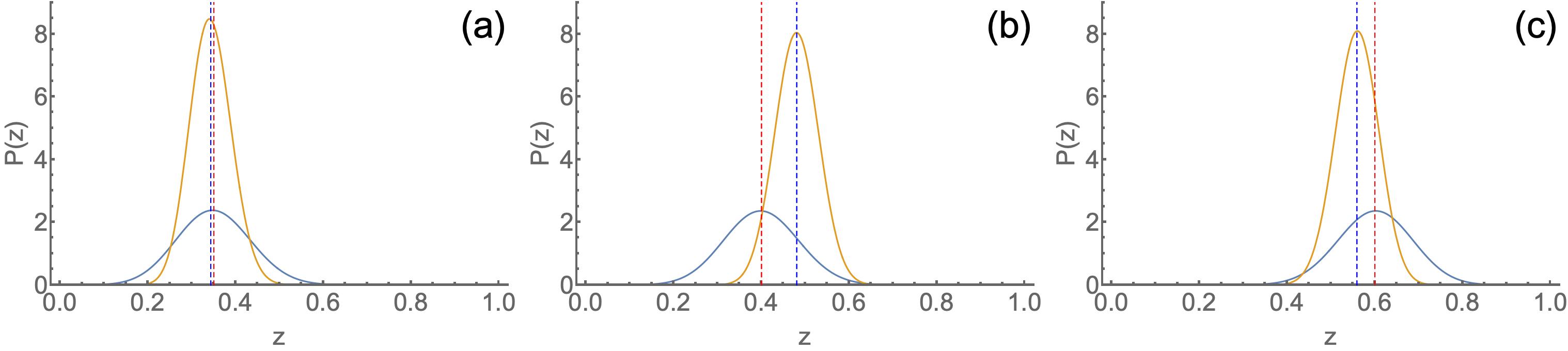}
\caption{\label{fig:qci100x3UP} For three simulations, (a), (b), and (c), of 100 interactions by expectation sampling between an $N=4$ quantum agent, Quinn, who takes a SIC measurement and an $N=2$ classical agent, Clark, who takes the equivalent of a Pauli Z measurement: The blue curve is the marginal $z$ distribution of Quinn's final posterior and the orange curve is Clark's final posterior. Both agents have uniform initial priors over their domains: Quinn's initial prior is uniform over the Bloch ball and Clark's initial prior is uniform over the unit interval. The blue dashed line is the mean of Clark's final posterior and the red dashed line is the mean of the marginal $z$ distribution of Quinn's final posterior. Quinn's posterior appears more spread because the $x$ and $y$ components of her posteriors are irrelevant to $z$. For all three, the posterior means vary less and grow closer as interactions increase; the agents seem to be coming to agreement to the extent possible for the less capable classical agent.}
\end{figure}

\begin{figure}[H]
\centering
\includegraphics[width=1.0\textwidth]{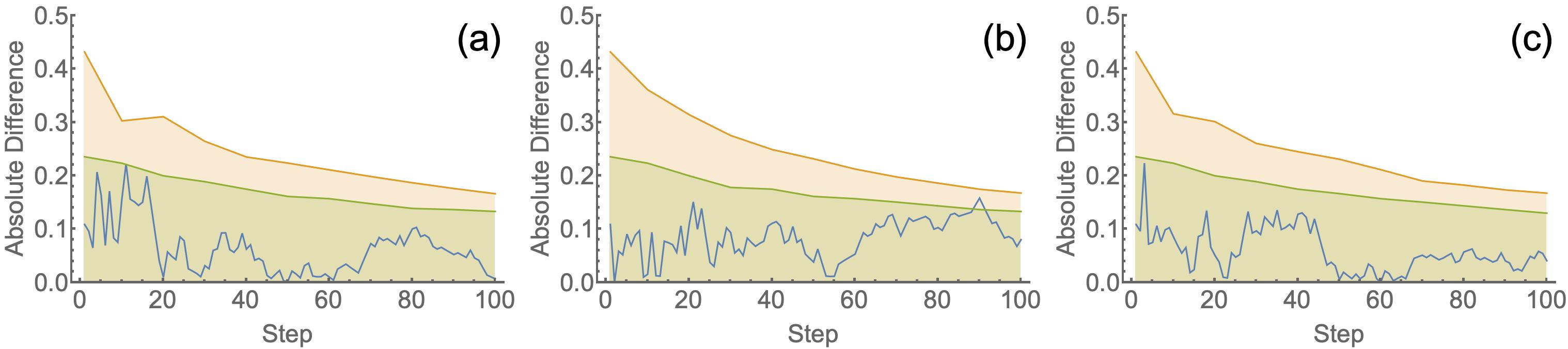}
\caption{\label{fig:qci100x3UP-meandiff} For the simulations (a), (b), and (c) of 100 interactions by expectation sampling between an $N=4$ quantum agent, Quinn, who takes SIC measurements and an $N=2$ classical agent, Clark, who takes the equivalent of a Pauli Z measurement: The blue curve tracks the trace distance between Clark's posterior mean and the mean of the marginal $z$ distribution of Quinn's posterior. The green curve is the standard deviation of Clark's posterior, computed every 10 interactions. The orange curve is the standard deviation in the $z$ direction of Quinn's posterior, computed every 10 interactions. The trend towards agreement in the $z$ direction is clear. The standard deviation in the $z$ direction of Quinn's posterior is larger because the $x$ and $y$ components of her posterior are irrelevant to $z$.}
\end{figure}

Suppose Quinn and Clark both have uniform initial priors over their physically valid regions. For Quinn, this is the Bloch ball, and for Clark, this is the chord in the Bloch ball aligned with the $z$ axis. Figure \ref{fig:qci100x3UP} displays the results of three simulations of 100 interactions between the $N=4$ quantum agent, Quinn, and the $N=2$ classical agent, Clark. Quinn always takes the the $d=2$ SIC reference action and Clark always takes his reference action, corresponding to the Pauli Z measurement. We plot the marginal $z$ distribution of Quinn's final posterior alongside Clark's final posterior. Figure \ref{fig:qci100x3UP-meandiff} displays the absolute difference between the $z$ component of Quinn's posterior means and Clark's posterior means after each interaction and compares them to the $z$ standard deviation of each. The results suggest that agreement in the $z$ dimension, the only direction for which Clark has any beliefs, should be expected.

The other kind of classical agent we consider is one with the \emph{same} reference action cardinality as the quantum agent, namely $N=d^2$. Suppose Clara is such an agent. Rather than being unaware of quantum mechanics like Clark, Clara doesn't \emph{believe} in it. One can imagine Clara having access to an apparatus designed by someone else to enact a SIC reference action, using it to assign conditional probability matrices to other available apparatuses, and then insisting on the classical physical postulate to relate the probabilities she would assign to the SIC measurement to those of whichever actual measurement she plans to perform. Clara reasons that the SIC can function as a reference measurement because she takes the classical view that reference probabilities surely represent knowledge of which SIC outcome is physically present, whether or not a SIC measurement is performed. 

%A necessary condition for a expectation sampling interaction to be well defined is that $N$, the number of outcomes of the reference action, for both agents be the same. The classical agents we considered in \S\ref{sec:exogenous} and \S\ref{sec:bernoulli} had $N=2$; there is no quantum counterpart for a reference action with this many outcomes. The smallest type of quantum system, where the Hilbert space dimension is $d=2$, corresponds to a reference action with $N=4$ outcomes. The agents in \S\ref{sec:2qubit} were agents of this sort. Thus, to consider the simplest case of a quantum and a classical agent interacting, we need two agents with $N=4$.

How can we arrange for Quinn and Clara to interact? As the reference simplex of each agent is the same, arranging for this type of classical agent to interact with a quantum agent simply requires that their physically valid regions are compatible. In order to ensure that Clara's mean parameter value is within Quinn's physically valid region, we restrict the support of Clara's prior to coincide with quantum state space in the SIC representation. If the mean value of Clara's prior density were outside this set, Quinn's physical postulate would not produce a valid probability distribution for some actions. Conceptually, then, this is an interaction between a quantum agent and an epistemically restricted classical agent. As with any such restriction of the prior support, we expect some of Clara's posteriors to display abrupt truncation at the boundary (Recall the case of disjoint priors in \S\ref{sec:bernoulli}, Figure \ref{fig:disjoint}.). 

Let both Quinn and Clara have uniform initial priors over the Bloch ball region. This time, Quinn may take one of the three Pauli actions, \eqref{paulix}, \eqref{pauliy}, and \eqref{pauliz}, for each interaction. She has a uniform utility function for the Pauli outcomes, such that, for each interaction, she chooses between the three Pauli measurements with equal probability. As any $m\times 4$ stochastic matrix is a physically valid conditional probability matrix for an $N=4$ classical agent, we can choose any we like for Clara's actions. In particular, the conditional probability matrices that Quinn uses for the Pauli actions are valid for Clara as well. We first consider the case where in each interaction Clara makes one of these measurements. Like Quinn, she has a uniform utility function and so each choice is randomized.

\begin{figure}[H]
\centering
\includegraphics[width=\textwidth]{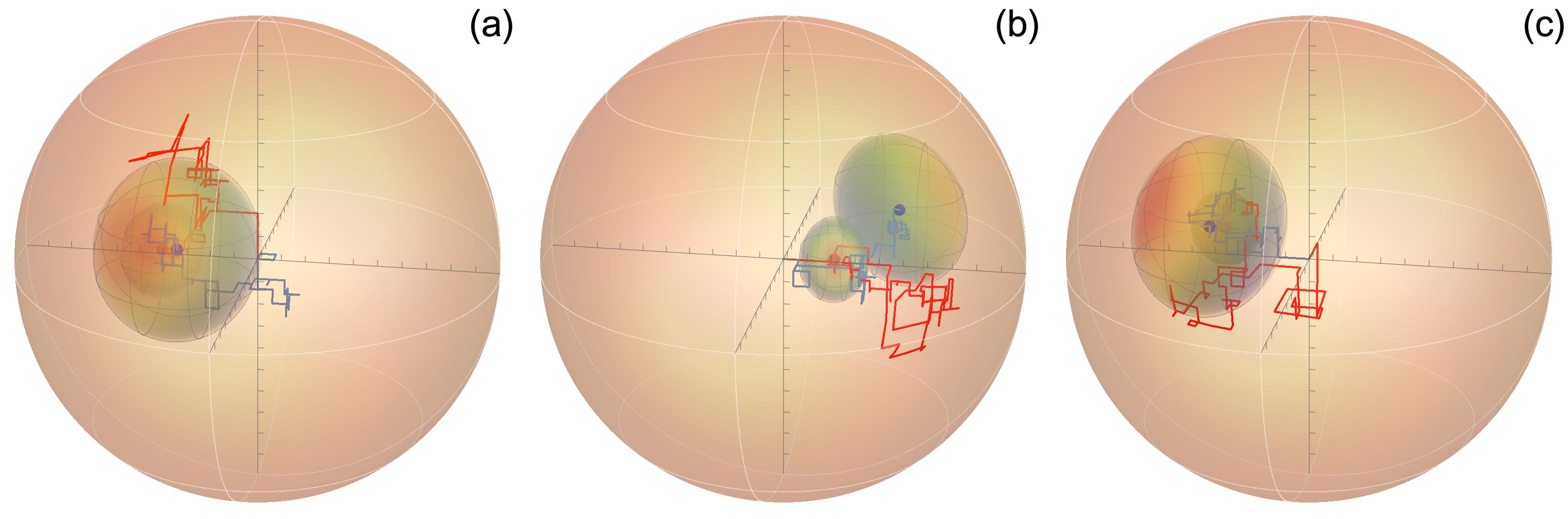}
\caption{\label{fig:qc100x3UP} Bloch ball plots of the final posterior densities within the standard deviation ellipsoids for three simulations, (a), (b), and (c), of 100 interactions by expectation sampling between an $N=4$ quantum agent, Quinn, and an $N=4$ classical agent, Clara, who each take one of the three Pauli measurements and both have uniform utility functions. Both agents have uniform initial priors on the Bloch ball region; for Clara, this is a proper subset of the physically valid region. The red dot is the mean of Quinn's final posterior and the blue dot is the mean of the Clara's final posterior. The red and blue lines track their previous posterior means. In each simulation, both agents end up with peaked posteriors, but Clara's final posteriors are more spread and display a higher probability density towards the boundary of the Bloch ball.}
\end{figure}

\begin{figure}[H]
\centering
\includegraphics[width=1.0\textwidth]{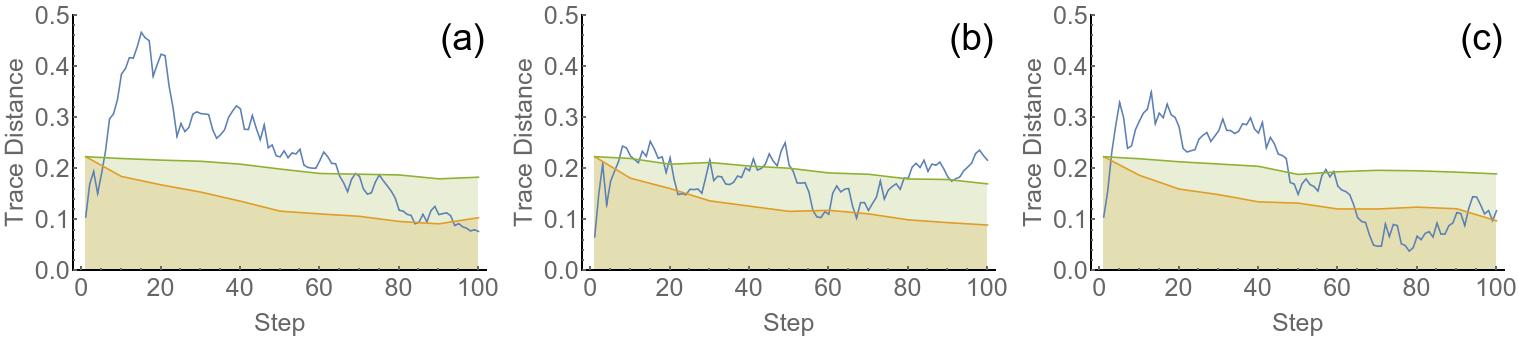}
\caption{\label{fig:qc100x3UP-trdist} For the simulations (a), (b), and (c) of 100 interactions by expectation sampling between an $N=4$ quantum agent, Quinn, and an $N=4$ classical agent, Clara, who each take one of the three Pauli measurements and have uniform utility functions and initial priors over the Bloch ball: The blue curve is the trace distance between the posterior means of the two agents. The orange curve is the length of the semi-major axis of the standard deviation ellipsoid for Quinn, computed every 10 interactions. The green curve is the length of the semi-major axis of the standard deviation ellipsoid for Clara, computed every 10 interactions. Clara's posteriors are more spread than the quantum agent's posteriors. For (a) and (c), the posterior means vary less and seem to be growing closer as interactions increase. It is less clear in (b).}
\end{figure}

Figure \ref{fig:qc100x3UP} contains density plots of the final posteriors within the standard deviation ellipsoids for three simulations of 100 interactions between Quinn and Clara. Because of her physical postulate, the actions Clara takes are less informative than Quinn's despite having the same conditional probability matrices. This leads Clara's posterior density to be significantly less peaked than Quinn's after the same number of interactions. Clara's lack of belief in quantum mechanics leads her to less certain conclusions in this case. One might say that, unlike Clark, Clara is limited by her choice of theory rather than her experimental capabilities. Figure \ref{fig:qc100x3UP-trdist} plots the trace distance between the means after each interaction and compares them to the semi-major axis of the standard deviation ellipsoids. The means appear to be approaching one another, although as Clara's densities are are so disperse, this trend may not be very meaningful.

Can Clara's theoretical limitations be offset if she has a different set of available actions? Unburdened by quantum theory and the associated demands of positive semidefiniteness, any stochastic matrix could be the conditional probability matrix for an action she might consider. Indeed, since Clara regards the reference action to be classical, the Pauli actions are, from her perspective, rather coarse measurements. Supposing Clara believes all of her uncertainty derives from her lack of knowledge of the true, preexisting reference measurement outcome, she might believe her possible actions correspond to perfectly sharp measurements.

Accordingly, our final example is a small modification of the previous one --- we only change the choice of actions for Clara. We choose conditional probability matrices for Clara which conceptually correspond to the SIC representation of the Pauli actions, but which are not consistent with any actual quantum measurement, meaning these matrices do not correspond to a reference action and POVM pair. Specifically, the conditional probability matrices corresponding to Clara's actions this time are 
\begin{equation}\label{cpaulis}
    X=\begin{pmatrix}
    1&0&1&0\\
    0&1&0&1
    \end{pmatrix}\;,\quad
    Y=\begin{pmatrix}
    1&0&0&1\\
    0&1&1&0
    \end{pmatrix}\;,\quad {\rm and}\quad 
    Z=\begin{pmatrix}
    1&1&0&0\\
    0&0&1&1
    \end{pmatrix}\;.
\end{equation}
The sense in which these conceptually correspond to the Pauli's is borrowed from \cite{DeBrota2020b}. There, the image of the probabilities and conditional probabilities of a physical postulate under the principal square root of the $\Phi$ matrix introduced in \S\ref{sec:postulates} corresponds to discrete quasiprobability representations of quantum mechanics. In the $d=2$ SIC representation, the matrices \eqref{cpaulis} are obtained from $R_X\Phi^{1/2}$, $R_Y\Phi^{1/2}$, and $R_Z\Phi^{1/2}$, respectively, where the $R$ matrices are the conditional probability matrices for the Pauli actions, equations \eqref{paulix}, \eqref{pauliy}, and \eqref{pauliz}. For an arbitrary action, right multiplying the conditional probability matrix by $\Phi^{1/2}$ would produce negativity; the Pauli actions happen to be appropriately oriented to avoid this with respect to the $d=2$ SIC we have chosen.

\begin{figure}[H]
\centering
\includegraphics[width=\textwidth]{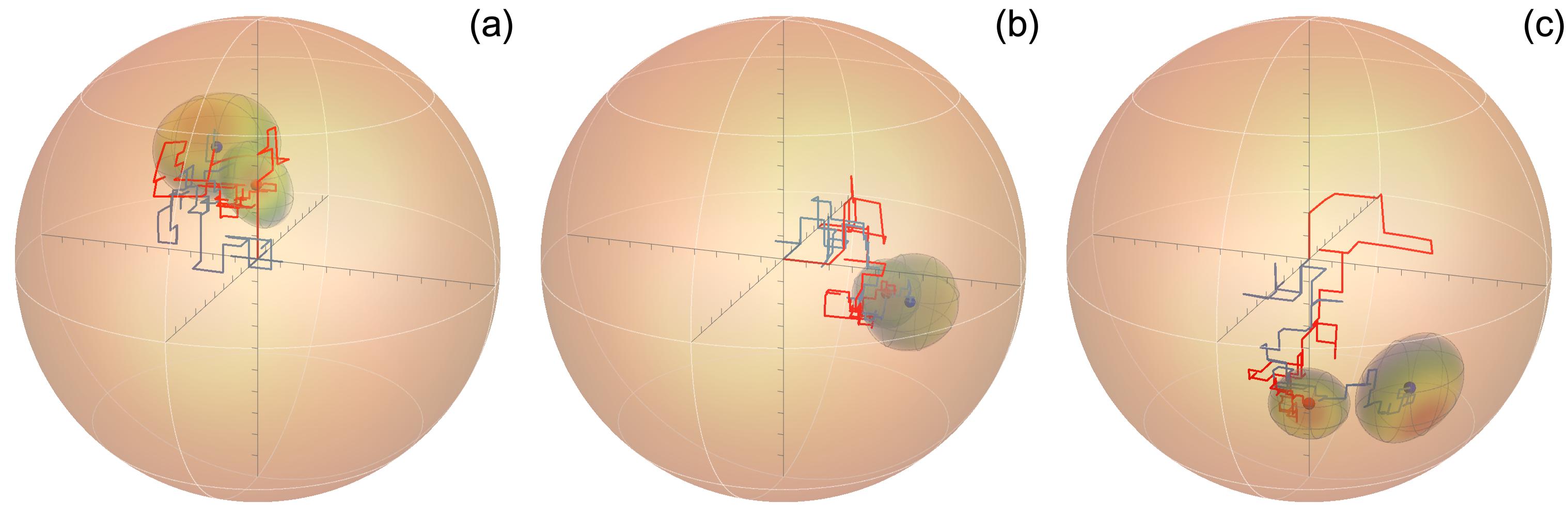}
\caption{\label{fig:qcCPaulis} Bloch ball plots of the final posterior densities within the standard deviation ellipsoids for three simulations, (a), (b), and (c), of 100 interactions by expectation sampling between an $N=4$ quantum agent, Quinn, who takes one of the three Pauli measurements and an $N=4$ classical agent, Clara, who takes one of the three measurements \eqref{cpaulis} and who both have uniform utility functions over the six outcomes of their possible actions. Both agents have uniform initial priors on the Bloch ball region; for Clara, this is a proper subset of the physically valid region. The red dot is the mean of Quinn's final posterior and the blue dot is the mean of Clara's final posterior. The red and blue lines track their previous posterior means. For each simulation, the final posterior means are in similar regions of the Bloch ball, suggesting agreement. Clara's final posteriors are more peaked than they were in Figure \ref{fig:qc100x3UP}, where she instead made a Pauli measurement, but compared to Quinn in these simulations, Clara's final posteriors are still more spread and display a higher probability density towards the boundary of the Bloch ball.}
\end{figure}

\begin{figure}[H]
\centering
\includegraphics[width=1.0\textwidth]{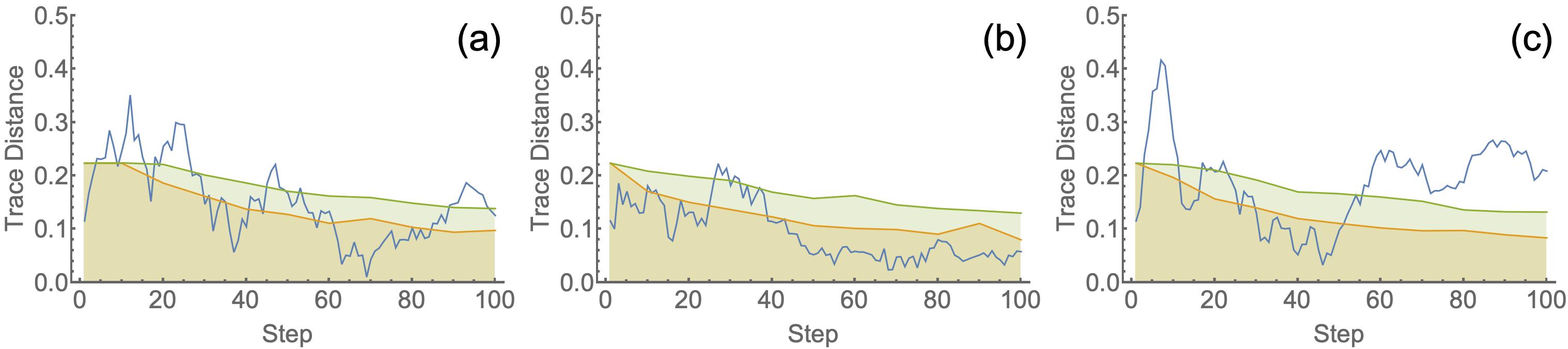}
\caption{\label{fig:qcCPaulis-trdist} For the simulations (a), (b), and (c) of 100 interactions by expectation sampling between an $N=4$ quantum agent, Quinn, who takes one of the three Pauli measurements and an $N=4$ classical agent, Clara, who takes one of the three measurements \eqref{cpaulis} and who both have uniform utility functions and initial priors over the Bloch ball: The blue curve is the trace distance between the posterior means of the two agents. The orange curve is the length of the semi-major axis of the standard deviation ellipsoid for Quinn's posteriors, computed every 10 interactions. The green curve is the length of the semi-major axis of the standard deviation ellipsoid for Clara's posteriors, computed every 10 interactions. Clara's posteriors are more spread than Quinn's posteriors, but the semi-major axis shrinks faster than when both quantum and classical agents take actions with the same conditional probability matrices (compare to Figure \ref{fig:qc100x3UP-trdist}). For (a) and (b), the posterior means seem to be growing closer as interactions increase. It is less clear for (c). }
\end{figure}

Figure \ref{fig:qcCPaulis} contains density plots for three simulations of 100 interactions between Quinn and Clara. Compared to the previous simulations, Clara's final posteriors are less spread; the reason for this is that her actions in this case are sharp, that is, no additional distribution flattening is introduced by the conditional probabilities. Figure \ref{fig:qcCPaulis-trdist} plots the trace distance between the means after each interaction. As before, there appears to be some trend towards agreement, but with only 100 interactions, it is hard to make a good judgement. 

\section{Discussion}\label{sec:disc}

In this final section, we provide clarifications and context, address some objections, and discuss possible future modifications and extensions of our work. 

We have developed an agent-based treatment of the classical and quantum multi-agent settings based on QBism. To accomplish this, we added a physical postulate on top of the rational agent framework from Bayesian decision theory. The secret ingredient is the idea of a reference action from QBism; with a physical postulate, an agent's expectations for a reference action coherently fixes their expectations for any other action. After understanding what classical and quantum agents are, we plunged them into two physically important scenarios: an agent receiving systems from an exogenous source and two agents interacting with one another. In the former, we saw that our agents can perform state tomography, arriving at a sharp estimate for the state prepared by the source. In the latter, we saw that interaction by expectation sampling can in some cases lead to agreement between two agents.

In all of our examples, we restricted attention to the case of exchangeable joint reference priors. When interacting with an exogenous source, this case is of immediate significance as it corresponds to an agent's belief that they are repeating \emph{the same} experiment in succession. It was also useful to work in this regime when defining an interaction, as interacting with an exogenous source is a special case of two agents interacting by the expectation sampling scheme proposed in \S\ref{sec:priorsampling}. In addition to being more tractable than fully general reference priors, there is much that is still not known about this special case.

One drawback of restricting to exchangeable priors is that there cannot be any entanglement between the systems of an exchangeable joint quantum state. Note that this does not prevent entanglement ``within'' each system; we only considered two-level quantum systems, but one could consider an exchangeable joint state where each system has a composite dimension Hilbert space such that some values of the parameter in the de Finetti representation contain entanglement. This way, there could be an exogenous source which prepares, for example, an entangled pair of qubits and an agent could have an exchangeable joint reference prior for qubit signals received two at a time. 

It's also, of course, possible to work directly with joint reference priors which aren't exchangeable in any way; one challenge is to identify a suitable extension or replacement for expectation sampling in this case. We will devote a sequel paper to addressing entanglement in our model. This, along with a treatment of coherence-preserving operations, is especially important if we want our agents to be able to enact quantum information processing tasks. 

From a slightly more realistic standpoint on agents, if an agent \emph{knows} they are interacting with another agent, rather than a source they believe will prepare the same state every time, it seems unlikely that they would regard the systems they receive exchangeably, as they might expect another agents' beliefs will change in response to data as their own do. If exchangeability is unwarranted, this possibility raises the question of whether there is any simplification of a joint prior which would reasonably apply in the case where two agents are interacting. In a similar vein, one might additionally speculate that some aspect of the breakdown in the usefulness of exchangeable priors could function as an \emph{indication} of agency.

Bayesian rational decision-makers are at the root of our approach. We did not modify this foundation in any way --- only added to it the ideas of a physical postulate and a way for agents to interact. How strong is the foundation we have chosen? Are there other choices? 

The so-called expected utility hypothesis has been used since the mid-twentieth century in economics to model consumer behavior. The idea is that humans are close enough to rational agents, suggestively named \emph{Homo economicus}, that this model will function as a useful large-scale approximation. Unfortunately for this intuition, humans behavior is typically quite far from this particular meaning of rational, a point most famously made by the Allais paradox~\cite{Allais1953}. Not only do we generally lack the resources to achieve coherence, the structure of actual human preferences in many cases renders the norm of coherence inapplicable. These considerations have spurred the development of different paradigms for capturing what is actually seen. These include bounded rationality \cite{Simon1997}, nudge theory \cite{Thaler2009}, and, perhaps the most well-known, prospect theory \cite{Kahneman1979} and cumulative prospect theory~\cite{Tversky1992}. 

As mentioned in \S\ref{sec:rationalagents}, the reason we chose the rational agent paradigm is that it provides us with simple decision makers who correctly use probability theory. We do not aspire to model actual behavior, human or otherwise. We set out to model conceptually idealized actors capable of using quantum theory through their ability to use probability theory. One does not need to be human to use probability theory or quantum theory correctly. In fact, it might be a disadvantage. 

Quantum theory is couched in probabilities. If the subjective Bayesians are correct concerning the meaning of probabilities, then the QBists are correct that quantum theory as currently understood is \emph{for} those who strive for this kind of rationality. We do not have the tools to do more than speculate at this time, but one wonders the following: If quantum theory as we presently conceive it is something which can be adopted by agents adhering to \emph{this} meaning of rationality, might it also have a representative with respect to a different one, perhaps where probabilities are not used?

Another aspect of our approach that takes place before anything physical is introduced is probability updating. In our development, it is assumed that an agent's new prior upon obtaining an outcome is the Bayes rule posterior implied by their prior and likelihood. It is often overlooked that this constitutes an assumption --- certainly, in many cases, a rational agent's new probabilities may be equal to their previous posterior, but coherence alone does not enforce any particular relation between different times. In fact, there are proposals which seek to extend ``rationality'' across time. The key to \emph{diachronic} coherence is allowing that an agent's beliefs about their own future beliefs may coherently constrain their current ones and vice versa. One particularly promising result within this scope is van Fraassen's reflection principle \cite{vanFraassen1984}. Revision of beliefs by Bayes rule as well as the more general Jeffrey's probability kinematics are seen to be special cases of reflection in the right circumstances \cite{Skyrms1987}. Then, if diachronic coherence is accepted as norm of rationality, an agent might choose to adopt one or another probability updating strategy in order to avoid a possible incoherence with their own preferences. In the examples we considered in \S\ref{sec:inference} and \S\ref{sec:interaction}, the beliefs of our agents \emph{do} change by Bayes rule updating, but other updating procedures may be considered in principle.

The next element of our approach is the idea that an agent is made ``quantum'' or ``classical'' by the physical postulate they adopt to constrain their expectations for what they regard to be a reference action to those of any other action. As we noted in \S\ref{sec:refactions}, this perspective is often advocated in expositions of QBism. Conceptual criticisms of QBism abound, but the \emph{technical} soundness of our development here is not at stake; nothing in this paper hinges on the validity of QBism as an interpretation. On the other hand, QBism's motivational role was indispensable.

Now we turn to the concept of interaction between agents. Here there were many ways to proceed. Our examples demonstrate that the example we chose, expectation sampling, is an interesting choice, but it is far from the only choice one could make. We expect there are many alternative and also interesting options. 

The most immediate alternative to expectation sampling might be called \emph{prior} sampling. This interaction is the same as expectation sampling, but instead of sampling the \emph{expectation} of another agent's prior, outcomes could be distributed by sampling a \emph{sample} of the other agent's prior. Depending on one's simulation purposes, prior sampling may be preferred over expectation sampling. 

As with expectation sampling, some regularizations will be needed in a prior sampling setting as well. The following two examples are due to Chris Ferrie. Suppose Alice and Bob are classical coin flipping agents interacting by prior sampling and both have the initial prior density
\begin{equation}
    P(\theta)=\frac{1}{2}\delta(\theta-0)+\frac{1}{2}\delta(\theta-1)\;.
\end{equation}
For this prior, prior sampling results in the preparation of a 0 or a 1 biased coin, that is, either a two-sided tails coin or a two-sided heads coin. If, as with expectation sampling and our simulations, both agents prepare their coin for the other simultaneously, there is a 50\% chance that Alice and Bob prepare opposite coins for each other. After flipping these coins, they will completely disagree and, in fact, cannot even communicate again as each believes the outcome the simulation will give them for a futher interaction to be impossible. The other 50\% of the time they will agree perfectly after a single round. Interestingly, if we instead imagine the agents take turns sampling and flipping, agreement is guaranteed after a single round.

Taking turns is not sufficient to avoid the possibility of belief polarization, however. Suppose now that Alice and Bob are qubit quantum agents interacting by prior sampling and both have the initial prior density 
\begin{equation}
    P(\psi)=\frac{1}{4}\delta(\ket{\psi}-\ket{0})+\frac{1}{4}\delta(\ket{\psi}-\ket{1})+\frac{1}{4}\delta(\ket{\psi}-\ket{+})+\frac{1}{4}\delta(\ket{\psi}-\ket{-})\;,
\end{equation}
indicating they each believe with uniform probability that the other will prepare one of the eigenstates of $Z$ or $X$. If they take turns sampling and measuring $Z$ or $X$, it is not hard to see that prior sampling can lead to situations where their posteriors are completely disjoint distributions consisting of orthogonal states. 

Although expectation sampling and prior sampling are closely related to an agent's physical postulate, we wish to again be absolutely clear that interactions in general are not compelled by quantum or classical mechanics in any direct sense. Expectation sampling itself is motivated by the imagery of broadcasting and receiving signals that we discussed in \S\ref{sec:priorsampling}. In light of this and the way we chose to generate outcomes for agents from exogenous sources in \S\ref{sec:coin} and \S\ref{sec:tomography}, expectation sampling is certainly a natural construction. However, as we saw, regularization of some kind is necessary to ensure interactions are well-defined in some cases. Perhaps there are better regularizations than we considered or generalizations of expectation sampling which don't have this difficulty. One can also imagine variants where outcomes for all agents are generated from one postulate, irrespective of the particular postulates of the agents involved.

Agent-based simulations of Wigner's friend scenarios will require a more involved interaction choice. The reason is that in this case, rather than only sending and receiving signals, we wish for at least one agent to be capable of directly manipulating another agent as a quantum system. An interesting possibility to consider in a case where an agent can directly act on another agent is that we might arrange for the \emph{action} of one agent to affect the \emph{outcome} for the other.

The fact that we can consider a diversity of interaction choices makes it especially clear that agreement between agents is not physically guaranteed in our agent-based approach. Prior to anything physical entering the picture, this is akin to recognizing that rationality also does not guarantee agreement because coherence is about a single agent's \emph{internal} consistency and the beliefs and experiences of distinct agents need not coincide. Indeed, for us, outcomes for each agent in an interacting pair are explicitly private. Even in situations of shared evidence, agreement between Bayesian agents is not guaranteed~\cite{Nielsen2021}. This lesson carries over to the quantum setting as well, where even infinite data is sometimes insufficient to ``wash out'' priors~\cite{Fuchs2009}. 

When outcomes are private, a famous result of Aumann establishes that Bayesian agents cannot ``agree to disagree'' if their priors are equal and their expectations are common knowledge, meaning they each know the values, know that the other knows, and so on~\cite{Aumann1976}. This result was later extended into a setting similar to ours, where agents iteratively exchange information and update their posteriors~\cite{Geanakoplos1982}. There, the common knowledge criterion can be reached in a finite number of steps. Aaronson later extended this result further, taking into account communication cost, complexity, and approximate agreement~\cite{Aaronson2005}. There is likely some relevance to our topic in the vast amount of literature stemming from Aumann's original result, but we note several distinctions: we do not require common priors; our agents are not explicitly aware the others are Bayesian rational agents, or even agents at all; we do not assume a common sample space; and rather than exchanging information like expectations or posteriors, our agents merely receive sampled outcomes.  

Clearly intersubjective agreement is often possible, but one must be careful to understand when it should be expected and when it shouldn't; this is why the possibility of agreement guided our analysis in \S\ref{sec:interaction}. We remind the reader who feels intersubjective agreement should be the norm that, in many actual reasoning scenarios, several humans may together be acting as a single agent, while intersubjective agreement concerns agreement between distinct agents. Our approach does not address the preconditions for several entities to together constitute a single agent.

On the subject of agreement, it should be recognized that we only considered one sense in which two agents might agree. In \S\ref{sec:interaction}, we considered probabilistic agreement, where two agents agree when their reference probability distributions become similar due to interaction. Agents of this kind with the same physical postulate will come to agree about how likely certain outcomes of a given action are. Another sense of agreement might be agreement in a behavioral sense, meaning two agents generally come to \emph{act} in the same way as a result of interactions. For rational agents, coming to behavioral agreement by interacting would require that, due to their interactions, the same actions tend to maximize the expected utility for each agent. As we only considered agents with static utility functions, two agents with different utilities coming to behavioral agreement, if possible, would substantially differ from the probabilistic case. This line of thinking raises the intriguing possibility of considering interactions which actually cause a change in utility functions as well.

There are many more additional interesting future directions in which one could build upon our approach. Let's list a few of them.

Perhaps the most natural next thing to do is consider situations where there are more than two agents, some pairs of which can interact. To do this, we will need to specify an interaction network. The first thing to address is the structure of belief dynamics in a social network of classical agents interacting by expectation sampling. It will be interesting to see how treating the analogous situation with quantum agents differs. It will also be interesting to consider larger tuples of interacting agents, for which expectation sampling will need extension or revision. 

A significant motivation for considering an agent-based approach in the first place was the possibility that a clear and consistent treatment would provide a new perspective on quantum algorithm design. Towards this, one should first try to implement known algorithms. To make these feasible, it will be necessary to discuss entanglement. 

An interacting agent setting also opens the doors to consider agents playing games. As in algorithm design, our approach or some variation on it may provide a fresh perspective for quantum game theory. We expect there are interesting equilibria to investigate other than agreement; perhaps game theory will help to identify them. 

On the more foundational side, one might wish to consider physical postulates which are not classical or quantum. One family of foil theories can be reached by modifying the quantum expression \eqref{bornrule} by choosing a quasistochastic $\Phi$ which doesn't correspond to any quantum reference action. In this way, one could consider postulates ``between'' classical and quantum theory.

Another foundational question we could ask stems from the fact that, in an actual simulation, the number of actions an agent chooses between is finite. In this paper, we considered the physically valid region to be those reference probabilities which produce valid probabilities for \emph{any} measurement, regardless of the agent's capabilities. One could instead consider the region carved out by the agent's actual abilities. For a quantum agent, this would then be a larger region, containing quantum state space. It could be interesting to simulate agents with priors in these different regions and try to identify behavioral consequences.

%\begin{enumerate}
%        \item All our agents are ``classical''; what could it mean to treat agents themselves quantum mechanically? Quantum state encoding of entire rational agent?
%    \end{enumerate}

Despite the similar appearances of ``quantum'' and ``decision theory'', our approach is very different from Quantum Decision Theory (QDT)~\cite{Yukalov2008}. QDT is a replacement for the rational agent paradigm making use of the mathematics of quantum mechanics to introduce stochasticity into the decisions of an agent. QDT is a quantum mechanical model of decision making while our interest is in modeling decision makers who \emph{use} quantum theory. Our approach requires no modification of the original rational agent framework. 

Our work bears at least some motivational similarity to reference \cite{Khrennikov2016}, where the author attempts to bring together QBism and a theory of decision making. There, however, the main deliverable appears to be an alternative to the usual QBist approach. The concept of multiple interacting agents is also not addressed. 

In summary, we have taken the necessary first steps towards quantum and classical agent-based modeling. Simple simulations showcase just a fraction of the potential that these lines of thinking likely conceal. We hope this paper inspires others to join us in taking the next steps. 

\section*{Acknowledgements}
The authors thank Christopher Ferrie for his very thorough review and helpful comments on an earlier draft. JBD thanks Nick White for the weekly Bayesian reading sessions and helping with a key step in a technical argument. JBD also thanks David DeBrota and Jeff La for stimulating conversations and suggestions. The authors acknowledge support from FQXi. This work was supported by the NSF STAQ project (PHY-1818914).

%\bibliographystyle{abbrv}
%\bibliography{main}
\printbibliography
\appendix
\section{Agreement for coin flipping agents with uniform initial priors}\label{sec:proof}
In this appendix, we prove that after an arbitrary interaction history starting from uniform priors for both Alice and Bob, the Kolomogorov distance between their expected posteriors for another interaction is less than or equal to the corresponding distance between their priors. 

Let $K(p_1(x),p_2(x))$ denote the Kolmogorov distance measure, defined to be the supremum of the absolute value of the difference between the cumulative distribution functions,
\begin{equation}
   K(p_1(x),p_2(x))=\sup_{x}\abs{\int_0^x p_1(t)dt-\int_0^x p_2(t)dt}\;.
\end{equation}
The Beta distribution is defined by
\begin{equation}
    {\rm Beta}(x;\alpha,\beta)=\frac{x^{\alpha-1}(1-x)^{\beta-1}}{B(\alpha,\beta)}
\end{equation} 
where $B(\alpha,\beta)$ is the Beta function and $\alpha$ and $\beta$ are shape parameters. The uniform prior is a special case of the Beta distribution with shape parameters $\alpha=\beta=1$. The Beta distribution is the conjugate prior of the Bernoulli likelihood, so we can characterize the form of Alice and Bob's priors after an arbitrary interaction history, that is, after receiving an arbitrary number of 0s and 1s in $N$ interactions, explicitly as Beta distributions. Let $k$ be the number of 1s Alice has received and $l$ be the number of 1s Bob has received in $N$ interactions between them. Then their current priors are
\begin{equation}
    P_A(\theta)={\rm Beta}(\theta;k+1,N-k+1)\quad {\rm and} \quad P_B(\theta)={\rm Beta}(\theta;l+1,N-l+1)\;.
\end{equation}
The mean value of ${\rm Beta}(x;\alpha,\beta)$ is $\frac{\alpha}{\alpha+\beta}$, so on a subsequent interaction, Alice receives a 1 with probability $\frac{l+1}{N+2}$ and Bob receives a 1 with probability $\frac{k+1}{N+2}$. Thus Alice's expected posterior is
\begin{equation}
    {\rm ExpPos}_A(\theta)=\frac{l+1}{N+2}{\rm Beta}(\theta;k+2,N-k+1)+\frac{N-l+1}{N+2}{\rm Beta}(\theta;k+1,N-k+2)\;,
\end{equation}
and Bob's is the same with $k\leftrightarrow l$. We wish to compare the Kolmogorov distance between their priors and their expected posteriors. The Kolmogorov distance between two distributions is the supremum of the absolute value of the difference between their cumulative distribution functions. The cumulative distribution function of the Beta distribution is the regularized incomplete Beta function 
\begin{equation}
    I_x(\alpha,\beta)=\frac{1}{B(\alpha,\beta)}\int_0^x t^{\alpha-1}(1-t)^{\beta-1}dt\;.
\end{equation}
The cumulative distribution functions for the expected posteriors is then the weighted mixture of the appropriate regularized incomplete Beta functions. Thus we wish to show that 
\begin{equation}\label{kdist}
\begin{split}
   & \sup_\theta\abs{I_\theta(l+1,N-l+1)-I_\theta(k+1,N-k+1)}\geq \\
   &  \sup_\theta \biggr|\frac{k+1}{N+2}I_\theta(l+2,N-l+1)+\frac{N-k+1}{N+2}I_\theta(l+1,N-l+2)\\
   & -\left(\frac{l+1}{N+2}I_\theta(k+2,N-k+1)+\frac{N-l+1}{N+2}I_\theta(k+1,N-k+2)\right)\biggr|\;.
\end{split}
\end{equation}
Denote by $\Omega$ the argument of the absolute value on the left hand side. Because of the absolute values, both sides are unchanged by $k\leftrightarrow l$ and when $k=l$, both sides are 0, so we may, without loss of generality, assume $N\geq k>l\geq 0$. The regularized incomplete Beta function has the recurrence properties
\begin{equation}
    I_x(\alpha+1,\beta)=I_x(\alpha,\beta)-\frac{x^\alpha(1-x)^\beta}{\alpha B(\alpha,\beta)}\;,
\end{equation}
and
\begin{equation}
    I_x(\alpha,\beta+1)=I_x(\alpha,\beta)+\frac{x^\alpha(1-x)^\beta}{\beta B(\alpha,\beta)}\;.
\end{equation}
These may be used to rewrite \eqref{kdist} as
\begin{equation}\label{kdist2}
    \sup_\theta|\Omega|\geq \sup_\theta \abs{\Omega - (N+1)!(k-l)\left(\frac{\theta^{k+1}(1-\theta)^{N-k+1}}{(k+1)!(N-k+1)!}+\frac{\theta^{l+1}(1-\theta)^{N-l+1}}{(l+1)!(N-l+1)!}\right)}\;.
\end{equation}
For $k>l$, $\Omega\geq0$ and so is the term subtracted from it on the right hand side. Thus  it would be sufficient if we could show that the argument of the absolute value on the right hand side of \eqref{kdist2} is nonnegative for all $\theta$. For integer values of $\alpha$ and $\beta$, the regularized incomplete Beta function takes the form
\begin{equation}
    I_x(\alpha,\beta)=\sum_{j=\alpha}^{\alpha+\beta-1}\frac{(\alpha+\beta-1)!}{j!(\alpha+\beta-1-j)!}x^j(1-x)^{\alpha+\beta-1-j}\;.
\end{equation}
Using this identity, $\Omega$ becomes a finite sum and we can divide out $(N+1)!$ so our problem reduces to showing the following inequality:
\begin{equation}
  \chi\equiv\sum_{j=l+1}^k \left[ \frac{x^j (1-x)^{N+1-j}}{j!(N+1-j)!} \right]
  - (k-l) \left[ \frac{x^{l+1} (1-x)^{N+1-l}}{(l+1)!(N+1-l)!} + \frac{x^{k+1} (1-x)^{N+1-k}}{(k+1)!(N+1-k)!} \right] \geq 0
\end{equation}
for $x \in [0,1]$ and integers $0 \leq l < k \leq N$.
Noting that the latter term has a factor of $(k - l)$, and the sum has exactly $(k-l)$ summands, we can combine the expression all into one sum:
\begin{align}
  \chi &=
  \sum_{j=l+1}^k \left[ \frac{x^j (1-x)^{N+1-j}}{j!(n+1-j)!}
  - \frac{x^{l+1} (1-x)^{n+1-l}}{(l+1)!(N+1-l)!} - \frac{x^{k+1} (1-x)^{N+1-k}}{(k+1)!(N+1-k)!}
\right]\nonumber
\\
  &=
  \sum_{j=l+1}^k \left[ \frac{x^j (1-x)^{N+1-j}[(1 - x) + x]}{j!(N+1-j)!}
    - (1-x)\frac{x^{l+1} (1-x)^{N+1-(l+1)}}{(l+1)!(N+1-l)!} - x \frac{x^{k} (1-x)^{N+1-k}}{(k+1)!(N+1-k)!}
\right]\nonumber
\\
  &=
  \sum_{j=l+1}^k \Bigg\{ (1-x)\left[\frac{x^j (1-x)^{N+1-j}}{j!(N+1-j)!}
    - \frac{x^{l+1} (1-x)^{N+1-(l+1)}}{(l+1)!(N+1-l)!}\right]
    \notag\\& \phantom{(n+1)!\sum_{j=l+1}^k~~~}
    + x\left[\frac{x^j (1-x)^{N+1-j}}{j!(N+1-j)!}
    - \frac{x^{k} (1-x)^{N+1-k}}{(k+1)!(N+1-k)!} \right]
  \Bigg\}\nonumber
\\
  &=
  \sum_{j=l+1}^k \Bigg\{ (1-x) \frac{x^{l+1} (1-x)^{N+1-(l+1)}}{(l+1)!(N+1-l)!}
    \left[\left( \frac{x}{1-x} \right)^{j-(l+1)}\frac{(l+1)!(N+1-l)!}{j!(N+1-j)!}~
    - 1\right]
    \notag\\& \phantom{(n+1)!\sum_{j=l+1}^k~~~}
    + x\frac{x^{k} (1-x)^{N+1-k}}{(k+1)!(N+1-k)!}
    \left[\left( \frac{1-x}{x} \right)^{k-j} \frac{(k+1)!(N+1-k)!}{j!(N+1-j)!}
    - 1 \right]
  \Bigg\}\;.
  \end{align}
We now separate the sums again:
\begin{align}
  \chi
  &=
  (1-x) \frac{x^{l+1} (1-x)^{n+1-(l+1)}}{(l+1)!(N+1-l)!} \sum_{j=l+1}^k
    \left[\left( \frac{x}{1-x} \right)^{j-(l+1)}\frac{(l+1)!(N+1-l)!}{j!(N+1-j)!}~
    - 1\right]
    \notag\\&~~
    + x\frac{x^{k} (1-x)^{N+1-k}}{(k+1)!(N+1-k)!}
   \sum_{j=l+1}^k
    \left[\left( \frac{1-x}{x} \right)^{k-j} \frac{(k+1)!(N+1-k)!}{j!(N+1-j)!}
    - 1 \right]
  .
\end{align}
Because $x \in [0, 1]$, the coefficient multiplying each sum on the left is nonnegative. So it will be sufficient for us to prove that each of the sums individually is nonnegative.
Furthermore, $x/(1-x)$ is monotonically increasing and $(1-x)/x$ is monotonically decreasing. Hence it is sufficient to prove nonnegativity of the first sum at $x=0$ and the second sum at $x=1$.

The first sum:
\begin{align}
  &
  \sum_{j=l+1}^k \left[\left( \frac{x}{1-x} \right)^{j-(l+1)}\frac{(l+1)!(n+1-l)!}{j!(n+1-j)!}~
  - 1\right]_{x=0}
  = \left[\frac{(l+1)!(n+1-l)!}{j!(n+1-j)!}\right]_{j = l+1} - (k-l)\nonumber
  \\
  =&
  (n+1-l) - (k-l) = n + 1 - k > 0
  .
\end{align}

The second sum:
\begin{align}
  &
   \sum_{j=l+1}^k
    \left[\left( \frac{1-x}{x} \right)^{k-j} \frac{(k+1)!(n+1-k)!}{j!(n+1-j)!}
    - 1 \right]_{x=1}
  = \left[\frac{(k+1)!(n+1-k)!}{j!(n+1-j)!}\right]_{j = k} - (k-l)\nonumber
  \\
  =&
  (k+1) - (k-l) = l + 1 > 0
  .
\end{align}

%Hence $\chi \geq 0$ and the result follows.

\end{document}